\pdfoutput=1
\documentclass{vldb}
\usepackage{graphicx}
\usepackage{balance}  % for  \balance command ON LAST PAGE  (only there!)

\usepackage{cite}
\usepackage{amsmath,amssymb,amsfonts}
\usepackage{algorithm2e}
\usepackage{textcomp}
\usepackage{xcolor}
\usepackage{array}
\usepackage{multirow}
\usepackage{url}
\usepackage{listings}
\usepackage{courier}
\def\BibTeX{{\rm B\kern-.05em{\sc i\kern-.025em b}\kern-.08em
    T\kern-.1667em\lower.7ex\hbox{E}\kern-.125emX}}
 
\definecolor{darkgreen}{rgb}{0.2,0.5,0}    
\newcommand{\mysys}{\textsf{sql4ml}\,}
\newcommand{\aka}{\emph{a.k.a.\,}}
\newcolumntype{P}[1]{>{\centering\arraybackslash}p{#1}}

\newcommand{\noop}[1]{}

\SetCommentSty{mycommfont}

\SetKwInOut{Parameter}{parameter}

\vldbTitle{sql4ml \\
A declarative end-to-end workflow for machine learning}
\vldbAuthors{Nantia Makrynioti, Ruy Ley-Wild, and Vasilis Vassalos}
\vldbDOI{https://doi.org/10.14778/xxxxxxx.xxxxxxx}
\vldbVolume{12}
\vldbNumber{xxx}
\vldbYear{2019}

\begin{document}

% ****************** TITLE ****************************************

\title{sql4ml \\
A declarative end-to-end workflow for machine learning}

% possible, but not really needed or used for PVLDB:
%\subtitle{[Extended Abstract]
%\titlenote{A full version of this paper is available as\textit{Author's Guide to Preparing ACM SIG Proceedings Using \LaTeX$2_\epsilon$\ and BibTeX} at \texttt{www.acm.org/eaddress.htm}}}

% ****************** AUTHORS **************************************


\numberofauthors{3} %  in this sample file, there are a *total*

\author{
% 1st. author
\alignauthor
Nantia Makrynioti\\
       \affaddr{Athens University of Economics and Business}\\
       \affaddr{Athens, Greece}\\
       \email{makriniotik@aueb.gr}
% 2nd. author
\alignauthor
Ruy Ley-Wild\\
       \affaddr{relationalAI}\\
       \affaddr{USA}\\
       \email{ruy.leywild@relational.ai}
% 3rd. author
\alignauthor Vasilis Vassalos\\
       \affaddr{Athens University of Economics and Business}\\
       \affaddr{Athens, Greece}\\
       \email{vassalos@aueb.gr}
}
% There's nothing stopping you putting the seventh, eighth, etc.
% author on the opening page (as the 'third row') but we ask,
% for aesthetic reasons that you place these 'additional authors'
% in the \additional authors block, viz.
%\additionalauthors{Additional authors: John Smith (The Th{\o}rv\"{a}ld Group, {\texttt{jsmith@affiliation.org}}), Julius P.~Kumquat
%(The \raggedright{Kumquat} Consortium, {\small \texttt{jpkumquat@consortium.net}}), and Ahmet Sacan (Drexel University, {\small \texttt{ahmetdevel@gmail.com}})}
% Just remember to make sure that the TOTAL number of authors
% is the number that will appear on the first page PLUS the
% number that will appear in the \additionalauthors section.

\maketitle

\begin{abstract}
We present \mysys, a system for expressing supervised machine learning (ML) models in SQL and automatically training them in TensorFlow. %
The primary motivation for this work stems from the observation that in many data science tasks there is a back-and-forth between
a relational database that stores the data and a machine learning framework. %
Data preprocessing and feature engineering typically happen in a database, whereas learning is usually executed in separate ML libraries. %
This fragmented workflow requires from the users to juggle between different programming paradigms and software systems. %
With \mysys the user can express both feature engineering and ML algorithms in SQL, while the system translates this code to an appropriate representation for training inside a machine learning framework.
We describe our translation method, present experimental results from applying it on three well-known ML algorithms and discuss the usability benefits from concentrating the entire workflow on the database side.
\end{abstract}

\section{Introduction}
\label{sec:intro}
Today data scientists suffer a fragmented workflow: typically feature engineering is performed in a relational database, while predictive models are developed with a machine learning library. %
Data scientists must tackle a fundamental impedance mismatch between relational databases and ML %
with the attendant accidental complexity of bridging the two worlds. %
The workflow involves %
transforming between data represented in relations and tensors (multi-dimensional arrays), %
context-switching between the declarative SQL and imperative Python (or R) programming paradigms, % knowing
and plumbing between siloed software systems. %
While relational databases provide advanced query optimization and permit selecting features by efficiently joining normalized tables, they are not always suited/optimal for linear algebra and have limited support for the type of iteration required for ML algorithms. %
In contrast, ML libraries provide mature support for linear algebra operations, automatic differentiation and mathematical optimization algorithms, although they are not adequate for relational queries and require the data matrix in a denormalized form.

\begin{figure}
  \includegraphics[width=\linewidth]{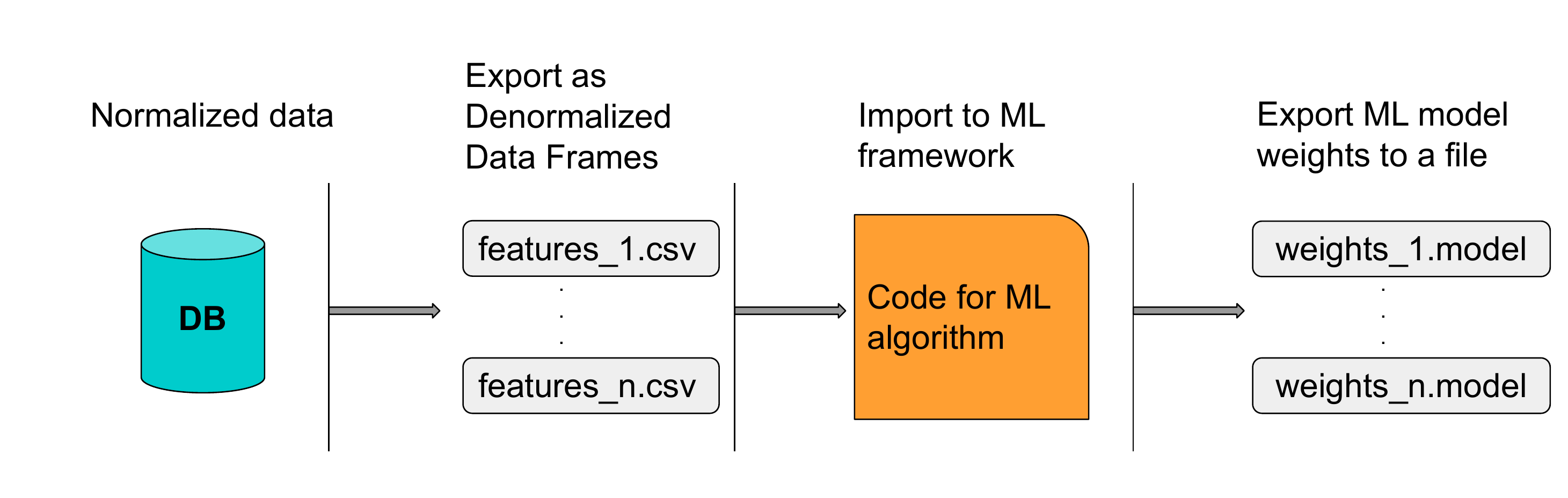}
  \caption{A typical data science workflow}
  \label{fig:ds_workflow}
\end{figure}

The overhead for interfacing databases and ML libraries materializes as increased costs: %
code increases in complexity and maintenance burden, %
projects need to specialize labor into data engineers on the SQL side and data scientists on the Python/R side, %
and the time to develop a model is slowed down by moving data through the pipeline and coordinating multiple people. %
The data science process (see Figure \ref{fig:ds_workflow}) is often experimental trying out different features, training sets, and ML models,
thus the overhead costs compound on every iteration.

We envision a more efficient workflow where an individual data scientist can perform both data management and ML in a single software system. % unified
We present \mysys as an integrated approach for data scientists to work entirely in SQL, while using a ML library like TensorFlow~\cite{abadi16} as the backend for ML tasks. %
We target users that are familiar with SQL and would like to have more advanced capabilities for data analysis. %
They need to understand the math for defining an ML model and write it in SQL, but do not need to deal with the lower level details of ML libraries, such as data batching and representation. %
Data representation in relations is standardized in the SQL language.

By synergizing a database and ML library, the user has more flexibility to define the algorithm of her choice compared to prepackaged ML algorithms. %
Our implementation specifically targets SQL and TensorFlow (Python API), although the ideas are equally applicable to other database languages such as Datalog, other ML libraries such as PyTorch, and other data science languages such as R. % any R equivalent popular library/framework?

In addition to using SQL for traditional data management, the user can also define and train supervised ML models in SQL. %
%\nm{what does this mean for unsupervised learning?}
She provides concrete SQL tables of features and target values (\aka labels) of training observations,
and specifies the ML model by SQL queries for the model prediction function and objective function in terms of an empty table of weights (\aka parameters or coefficients) which will be learned during training. %
% number of iterations
\mysys generates the boilerplate code to move the features from relations in the database to CSV files on disk or directly to tensors in the ML library. %

\mysys automatically translates SQL queries defining an ML model to TensorFlow Python code, which is sufficient to invoke gradient descent to find the optimal weights that minimize the objective function. %
% I think CRAN can do some of the things TensorFlow and PyTorch do.
The objective function written in SQL can be understood as a query for the optimal weights that minimize it. %
The generated TensorFlow/Python code is effectively a physical plan of the query. %
\mysys thus makes it easier for a single person to run the full data science cycle, because it eliminates the need for two programming paradigms and automates moving data between systems, thus leaving more resources to find the right features and model.

% models are developed in Python, R or MATLAB and other imperative languages. % and productionized into C++
% (SQL is a declarative relational algebra, wheras Python and R are imperative and sequential)

% avoid having to know py/tf
% avoid having to move data manually (programming burden)
% weights are stored in a custom format
% filesystem is a poor database
%
%\rlw{syntax-directed translation from SQL to Python}
%
%% data analysts use SQL
%% data scientists use Python
%
%\rlw{We seem to provide a SQL interface to learning in TF, where the user provides the prediction/objective functions.\\
%It would be interesting to provide a higher level SQL interface to learning a la scikit-learn with menu of readymade ML models.\\
%ML engineers program at the "sql2tf" layer, Data Scientists program at the "sql2sk" layer.
%}
%\nm{MADlib is a system for machine learning in Postgres and Greenpulm databases. %
%  It provides ready ML algorithms to call as user defined functions in SQL queries. %
%  Microsoft SQL Server provides a functionality to run Python or R code in stored procedures. %
%  BigQuery has a beta functionality similar to MADlib with a few ready models. %
%  I don't know how these models are implemented, maybe they use Python libraries like scikit-learn. %
%I think the main difference between these systems and our direction is that we try to be more flexible in letting the user define the objective function, hence not limiting her to a set of readymade models.}

% interoperability, combination, coordination

% store multiple model weights in the database instead of the filesystem;
Finally, \mysys moves the computed weights from the ML library back to the database. %
Thus, the data scientist can stay in the database to compute predictions on test data and statistics about the ML model, such as percent error or precision and recall. %
Furthermore, the database can be used to store multiple trained ML models and keep track of the features that were used in each ML model, in contrast to ad hoc management of weights in files on the filesystem.
In this way, she can also use the database as a system for keeping records of experimental results and manage the entire ML lifecycle in one place. %
% \rlw{branching might not be broadly known outside LB and isn't necessary to keep multiple models in a database, we can have a weights table with an extra column for the model name/run}

In summary, we provide an end-to-end workflow that starts in the database, pushes the training of the model to the ML library, and stores the computed weights back in the database. %
A core contribution is a method for translating objective functions of ML algorithms written in SQL to linear algebra operations in an ML library. %
We demonstrate how our translation method works and apply it on popular machine learning algorithms, including Linear Regression, Factorization Machines~\cite{Rendle10}, and Logistic Regression. %

The rest of the paper is organized as follows. %
Section~\ref{sec:overview} presents an overview of \mysys workflow in terms of input and generated code. %
In section \ref{sec:relational_vs_linear} we discuss differences in the functionality and data structures between relational databases and ML frameworks, %
whereas in section \ref{sec:bridge} we describe the steps of our approach in detail. %
Sections \ref{sec:benefits} and \ref{sec:experiments} present the usability benefits of \mysys and an experimental evaluation respectively. %
Finally, section \ref{sec:related} discusses related work and section \ref{sec:conclusion} concludes the paper.

% keep the figure at the end so it shows up on page 2

\section{Overview of sql4ml}
\label{sec:overview}
\begin{figure}
  \includegraphics[width=\linewidth]{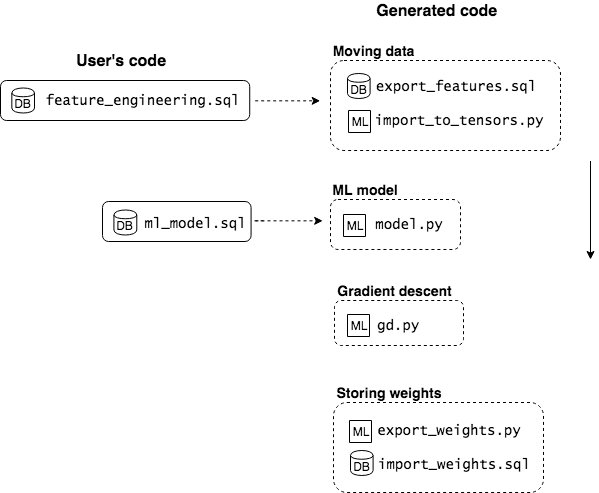}
  \caption{Overview of code involved in \mysys workflow}
  \label{fig:code_files}
\end{figure}

\mysys takes as input SQL code and generates Python code consisting either of TensorFlow API functions or other Python modules, %
which automates stages of the workflow in Figure \ref{fig:ds_workflow}. %
An overview of the code files involved in \mysys is presented in Figure \ref{fig:code_files} and highlighted next. %

\emph{User's code.} The user provides SQL code defining which features shall be used for training and %
the objective function of the ML model (left side of Figure \ref{fig:code_files}).

\emph{Generated code.} There are four main groups of code that \mysys generates, which correspond to the four dotted rectangles of Figure \ref{fig:code_files}. %
The first regards the transfer of data from relations to tensors. %
\mysys generates SQL queries for exporting feature and target values of training data and Python code for loading these values into tensors. %
In the second group, there is code for the definition of the ML model. %
This is generated by translating SQL queries that define the objective function of the ML model to TensorFlow API functions. %
Finally, boilerplate code is generated for triggering gradient descent and importing the computed values of weights back to database tables after training. %
The execution order of the generated code goes from top to bottom following the arrow in Figure \ref{fig:code_files}.

\emph{DB} and \emph{ML} icons in Figure \ref{fig:code_files} are used to indicate which parts of the code run on the database and which on the ML framework. %
This is also denoted by the file extensions \texttt{.sql} and \texttt{.py} on each filename (in \mysys we chose to translate SQL to Python code, including functions of the TensorFlow API, and use TensorFlow as an execution engine.) %
We describe in detail how each part of the code is generated and provide specific examples %
of user provided and automatically generated code in section \ref{sec:bridge}.

\section{On DB and ML foundations}
\label{sec:relational_vs_linear}
In this section we briefly discuss some fundamental differences between relational databases and ML frameworks, % 
which pose challenges in integrating functionality of the two worlds. %
Some of the differences stem from the foundations of relational algebra and linear algebra, whereas others are related to programming paradigms %
and the use of declarative versus procedural languages.

Relational databases are built around the abstraction of relation, which is an unordered set of tuples. %
There is no notion of order between the columns of a tuple. %
Because of this, the user is not concerned with the particular physical representation of relations in the database, and the system is free to optimize the representation (e.g., with indexes). %
On the other hand ML frameworks operate on tensors, which are multidimensional arrays. %
The word ``array'' is key here. %
It indicates that the elements of a tensor are accessed via indices, basically natural numbers from $0$ to $I-1$ (or $1..I$), which in turn implies a specific order on the elements. %
The connection between boolean matrices and relations is well-established, whereas recent work \cite{Brijder19} %
studies mappings between matrices of numeric values and $K$-relations. %

On the functionality front, relational operators map input relations to an output relation %
and include selection, projection, joins and set-theoretic operators. %
In contract, linear algebra operators include matrix multiplication, summations over rows and columns, matrix transpose, element-wise arithmetic %
operations and others. %
In addition to this and as far as programming model is concerned, databases support declarative %
languages, traditionally SQL and in some cases Datalog, whereas ML frameworks usually prefer %
imperative languages, such as Python and C++. %
It has been shown that a set of commonly used linear algebra operators can be implemented in SQL using element-wise mathematical operations and aggregations \cite{Kumar15}, \cite{Schleich16}, \cite{AboKhamis18}. %
For example, a matrix-vector multiplication is an element-wise multiplication between columns and a summation over the elements of each tuple in a relation. %
However, other parts of machine learning algorithms are more difficult to implement inside a database.

A significant limitation of SQL is the lack of iteration constructs. %
% \rlw{what kind of iteration? for a fixed number of iterations? until convergence?}
RDBMS support for iterations is typically limited to fixed-points over sets. %
Iterative processes are very common in machine learning though. %
The training of a ML model is one of them and is done using well-known mathematical optimization algorithms, such as gradient descent. %
Because of their wide use in machine learning, ML frameworks provide highly optimized implementations of such algorithms.

Another very useful capability of some ML frameworks, which is not currently supported in databases, is \emph{automatic differentiation}. %
Systems that do not provide automatic differentiation require the user to explicitly write the derivatives of the objective function, which becomes nontrivial for complicated functions.

On the other hand, ML systems lack support when it comes to data processing. %
Unlike databases, a tensor library like TensorFlow does not provide relational operators (e.g., joins and filters).
% Given that everything is stored in tensors it becomes difficult to combine information from different tensors or filter rows. %
% \rlw{(What's the difficulty? For example: addition combines two tensors, filtering can be done by masking with 0's. Is the issue e.g. with doing joins?)} %
% That is one of the main reasons why we prefer storing data in databases, instead of files. %
While a ML framework can be combined with other libraries in the host language providing tabular operations, such as Python Pandas \cite{pandas}, there are two main issues associated with this approach. %
First, the scalability of these solutions is limited by the main memory of the machine.
Second, such libraries offer a quite procedural way of writing relational operations that exposes the order in which they are executed, e.g. A.join(B).join(C).
On the other hand, databases incorporate an extensive body of work to optimize queries just describing the output set and handle large datasets.

\section{Bridging the gap}
\label{sec:bridge}
Given the mismatch between the data and programming paradigms in databases and machine learning systems, we propose a unified approach where both can be programmed from SQL. %
In \mysys, the data analyst writes SQL queries both to prepare the features and to define the ML task. %
In particular, the user writes the objective function of the ML model in SQL and the system automatically translates it to the corresponding TensorFlow code. %
Moreover, the data are transparently moved from relations in the database to tensors in the ML system.

% Given the data and programming model between databases and machine learning, we propose that users write the objective function of a machine learning model in SQL, then we provide an automatic translation to TensorFlow code and transparently move the data betwen relations and tensors. %
% When we refer to tensors in the context of our translation method, we mean either vectors or matrices.
% not quite writing machine learning algorithms in SQL

\subsection{Common components in ML algorithms}
A large class of supervised algorithms in ML, including linear regression, logistic regression, Factorization Machines and SVM, follow a common pattern in how they work. %
Unsupervised algorithms, such as k-means, can be reformulated to fit into this pattern, too, if we replace the concept of features with data points. %
We explain the encountered similarities below.	

\subsubsection{Prediction and Objective Functions}
Given a set of input features $x$, a machine learning algorithm can be specified by an objective function.
For example, in logistic regression the relationship between the target variable $h_\theta(x)$ and the input features $x$ is defined by the prediction function
\begin{equation}
\label{sigmoid_equation}
h_\theta(x)=\frac{1}{1+e^{-x^\top\theta}}
\end{equation}
%\begin{equation}
%\label{lr_equation}
%y_i = x_i^\top \beta,
%\end{equation}
%
where $\theta$ are the unknown weights of the ML model to be optimized. %
The objective function (\aka loss function or cost function) of logistic regression is
\begin{equation}
\label{logistic_loss}
-\frac{1}{n}\sum_{i}^{n}[y_i\log(h_\theta(x_i)) + (1-y_i)\log(1-h_\theta(x_i))]
\end{equation}

%\begin{equation}
%\label{error_equation}
% \frac{1}{n}\sum_{i}^{n}|y_i - \hat{y}_i|
% \quad\text{or}\quad
% \frac{1}{n}\sum_{i}^{n}(y_i - \hat{y}_i)^2.
%\end{equation}

%\rlw{maybe better for space to inline formulas in the paragraph?}

\subsubsection{Optimization}
The minimization/maximization of the objective function is done using a mathematical optimization algorithm, such as gradient descent. %
Gradient descent is an iterative method, which computes the derivative of the objective function and updates the ML model weights in each iteration. %
Given a system such as TensorFlow that supports automatic differentiation and implements a mathematical optimization algorithm, the user only needs to provide the objective function in terms of input features and weights.

%\rlw{not relevant to this paper?}
%In case of constraint optimization problems, constraints can be defined on the objective function and the weights, and the problem can be again optimized using a constraint optimization solver, such as linear or quadratic programming.
%\nm{TensorFlow does not provide linear or quadratic programming solvers in the core module. In principle though, one can rewrite a constrained optimization problem to an unconstrained by adding the constraints to the objective function. However, we do not provide an example for this in the paper.}

%In the next section, we describe the proposed architecture end to end and explain its advantages to the traditional data science workflow. %

\subsection{End-to-end workflow}
We propose a workflow, where the user writes SQL to both query the data and express the objective function of the ML model. %
A TensorFlow representation of the SQL queries defining the ML model and consisting of linear algebra operators and tensors is generated. %
%\nm{I think that a more generic phrasing would be more suitable to convey the message that our method is not limited to TensorFlow.}

%\rlw{move this to another section?}
%We assume that a ML library is well-optimized on linear algebra operators, so we attempt to take advantage of this fact when we recognize a linear algebra operation implemented in SQL. %
%For example, a relational sum of products of entries from two tables can be recognized to be a matrix multiplication. %
%Although the objective function is expressed in SQL, we do not evaluate it in the database and instead translate it to a tensor library to be optimized.

\begin{figure}
  \includegraphics[width=\linewidth]{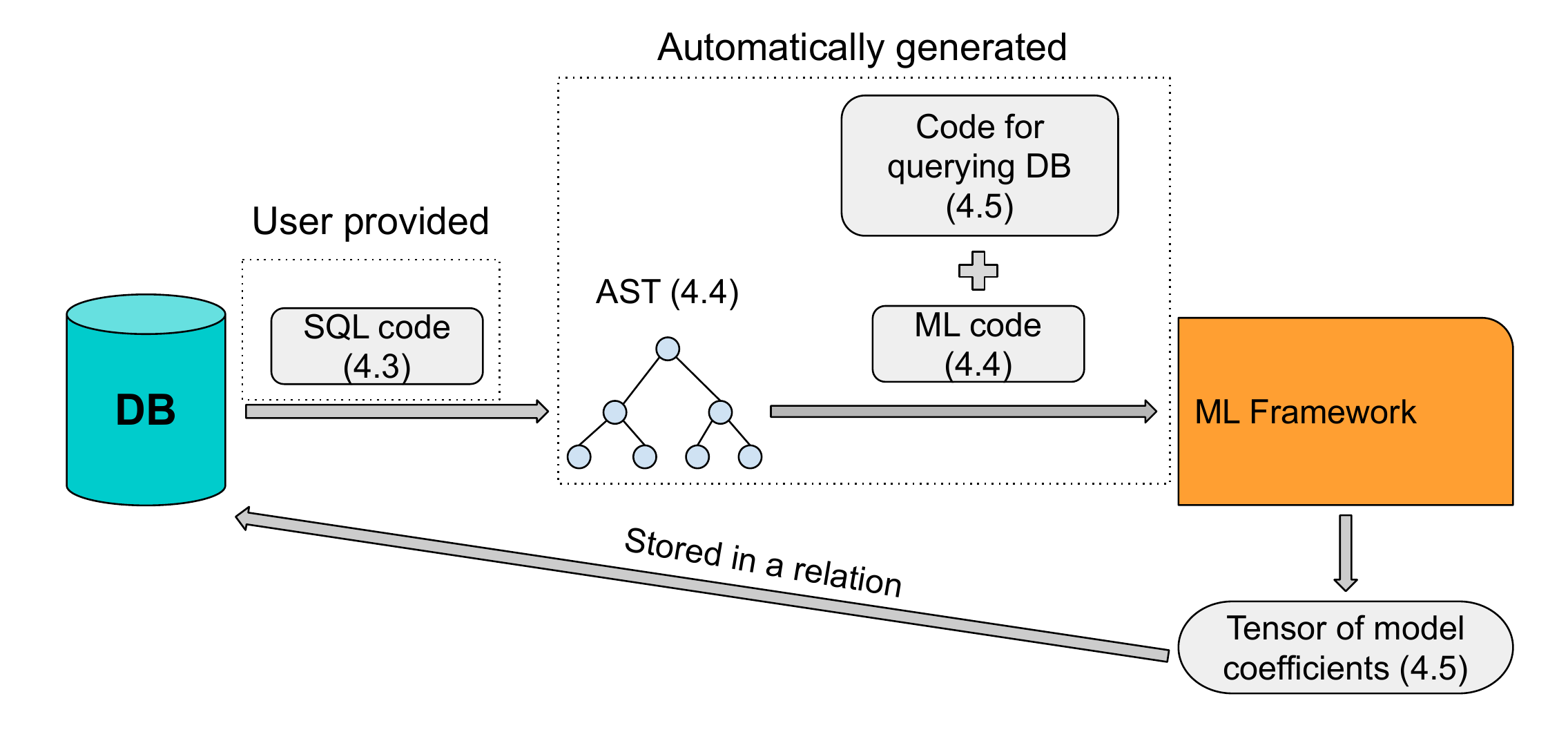}
  \caption{Proposed workflow between a DB and a ML system (numbers inside parenthesis indicate the subsection where each stage is described)}
  \label{fig:proposed_workflow}
\end{figure}

This representation is created in two steps. %
First, the SQL code defining the objective function of the ML model is converted to an abstract syntax tree (AST), which is then automatically translated to code in the host language of a ML framework.
%Then, the SQL code defining the objective function is converted to an abstract syntax tree, which is finally automatically translated to Python code consisting of functions of the TensorFlow API. %
This code is supplemented with a few more lines calling gradient descent on the objective function inside a loop. %
The entire program is then executed on the ML framework. %
%The entire program is then executed on the ML framework just like any other TensorFlow program. %
In order to feed data from relations to tensors, we automatically generate code either for connecting and querying the database using a suitable driver or for exporting data to files. %
After the generated code is executed and the optimal weights for the objective function are computed, they are transferred to relations inside the database. %
The proposed workflow is displayed in Figure \ref{fig:proposed_workflow}.

%The proposed workflow showcases three key advantages. %
%\begin{itemize}
%\item First, both the code for feature engineering and the code for defining the ML model are defined in a single declarative language, e.g. %
%SQL. %
%The user needs to understand the math operations of the objective function, but doesn't have to be familiar with the particular programming model of the ML system. %
%
%\item Second, there is no need to write ETL processes in order to move the data from one system to the other. %
%This is handled automatically.
%% and the data are read into memory. %
%% That's not true when we write data to files.
%
%\item Third, given that when the ML model is trained its weights are stored back to the database, one can handle the machine learning lifecycle inside the database storing multiple ML models and experiments, which can also be queried with SQL. %
%This eliminates the need for other systems providing tracking of experimental results and ML model versions.
%\end{itemize}

Note that joins needed in feature engineering are still evaluated inside the database, whereas any linear algebra operation written in SQL is translated to the appropriate operators of the ML framework and is evaluated in it. %
On ML frameworks supporting it, that also means the linear algebra part can run on GPUs. %
Relational databases provide quite advanced query optimization and there are already mature solutions for linear algebra operations, automatic differentiation and mathematical optimization algorithms outside the database ecosystem. %
In our approach the key components of a machine learning algorithm are provided using SQL, whereas the training of the ML model is executed on a system specializing in machine learning, hence avoiding the need for implementing an iterative process in the language of the database. %
%RDBMS support for iterations is typically limited fixed-points over sets. %
Essentially, we attempt to combine the best from both worlds, but at the same time do so in a transparent way, in order to reduce the manual work required by the user and the need to be familiar with lower level details of machine learning frameworks, such as data representation. %
Also by creating an under the hood synergy between a database and mathematical optimizers of ML frameworks, the user has more flexibility to define the algorithm of her choice compared to using ready made ML algorithms. %
%In this work we target TensorFlow and its Python API as an execution engine for the generated code.

\subsection{ML algorithms in SQL}
\label{sec:sql_examples}
In this section we present how machine learning algorithms are written in SQL through the example of logistic regression. %
%\nm{I'm not sure about Linear Regression. Maybe Logistic Regression is more suitable as an example that showcases what we do and is more complicated.}
%Other machine learning algorithms are included in the Appendix. %
As stated earlier, the user writes the objective function of the ML model in SQL. %
%\mysys automatically generates the Python code to train the ML model in TensorFlow.

We implement logistic regression on the following schema:

%\rlw{maybe pick a shorter name like rowId instead of observationId?}

\begin{lstlisting}[frame=single,label=schema]
observations(rowID: int)
features(rowID: int, featureName: string, 
v: double)
targets(rowID: int, v: double)
weights(featureName: string, v: double)
\end{lstlisting}

The \texttt{observations} table stores the identification numbers of training observations. %
For every observation and feature, the \texttt{features} table has an entry with the feature's value. %
For example, the entry \texttt{(1, "feature1", 30.5)} means that feature \texttt{"feature1"} of observation \texttt{1} has value \texttt{30.5}. %
The \texttt{targets} table stores the observed target values of the training observations. %
Data for these three tables are provided by the user, extensional to the database. %
Finally, the \texttt{weights} table is filled in by learning the weights of a model in TensorFlow.

Given these tables, we define SQL views \texttt{sigmoid} for the sigmoid function of the logistic regression model and \texttt{objective} for the objective function (see Listing \ref{logistic_loss_sql}). %
These functions embody the Equations \ref{sigmoid_equation} and \ref{logistic_loss} and are expressible in SQL using numeric and aggregation operations. %
Instead of evaluating them inside a database, they are translated to TensorFlow linear algebra functions for evaluation inside TensorFlow. %
Note that the user defines the model by writing prediction and objective functions in SQL, while \mysys generates the code to train the ML model.

\begin{lstlisting}[language=SQL,frame=single,caption=Logistic regression in SQL,label=logistic_loss_sql]
CREATE VIEW product AS 
  SELECT SUM(features.v * weights.v) AS v, 
  features.rowID AS rowID
  FROM features, weights
  WHERE features.featureName=weights.featureName
  GROUP BY rowID;

CREATE VIEW sigmoid AS 
  SELECT product.rowID AS rowID, 
  1/(1+EXP(-product.v)) AS v 
  FROM product;

CREATE VIEW log_sigmoid AS 
  SELECT sigmoid.rowID AS rowID, 
  LN(sigmoid.v) AS v 
  FROM sigmoid;

CREATE VIEW log_1_minus_sigmoid AS 
  SELECT sigmoid.rowID AS rowID, 
  LN(1-sigmoid.v) AS v 
  FROM sigmoid;

CREATE VIEW objective AS 
  SELECT (-1)*SUM((targets.v * log_sigmoid.v) + ((1-targets.v) * log_1_minus_sigmoid.v)) AS v 
  FROM targets, log_sigmoid, log_1_minus_sigmoid 
  WHERE targets.rowID=log_sigmoid.rowID and log_sigmoid.rowID=log_1_minus_sigmoid.rowID;
\end{lstlisting}

%\begin{lstlisting}[language=SQL,frame=single,caption=Linear regression in SQL,label=LR_sql]
%CREATE VIEW predictions AS
%    SELECT features.rowID AS rowID,
%           SUM(features.v * weights.v) AS v
%    FROM features, weights
%    WHERE features.featureName = weights.featureName
%    GROUP BY rowID;
%
%CREATE VIEW squaredErrors AS
%    SELECT errors.rowID AS rowID,
%           POW(predictions.v - targets.v, 2) AS v
%    FROM predictions;
%
%CREATE VIEW objective AS
%    SELECT AVG(squaredErrors.v)
%    FROM squaredErrors;
%
%\end{lstlisting}

%\rlw{-- FIX this nested query should work, but maybe it's more complicated to read so not worth it}
%\nm{Yes, I think it's harder to read.}    
%CREATE VIEW objective AS
%    SELECT AVG(squaredErrors.v)
%    FROM (SELECT errors.observationID              AS observationID,
%                 POW(predictions.v - targets.v, 2) AS v
%          FROM errors) qAS squaredErrors;

%\rlw{say something about how code is symbolic/unevaluated because it depends on \texttt{weights}}
%\nm{How about this?}
The SQL code as is cannot be evaluated because the \texttt{weights} table is empty. %
Even if the weights were initialized with random values, only a single round of evaluation would be possible. %
No iterative process is defined in the code of Listing \ref{logistic_loss_sql} to update tuples in \texttt{weights}. %
Using the SQL representation we produce a representation in TensorFlow/Python API to be evaluated in TensorFlow.

\subsection{From SQL to TensorFlow API}
In this section we describe the process of translating machine learning algorithms from SQL to TensorFlow API.
We start with a simple example, the prediction function of linear regression.
\begin{equation}
\label{lr_prediction}
y_i = x_i^\top \theta
\end{equation}
This may be implemented in SQL as follows:
\begin{lstlisting}[language=SQL,frame=single,label=prediction_sql]
CREATE VIEW predictions AS
  SELECT features.rowID AS rowID,
  SUM(features.v * weights.v) AS v
  FROM features, weights
  WHERE features.featureName = weights.featureName
  GROUP BY rowID;
\end{lstlisting}
%\rlw{maybe drop 'tf.' throughout?}
Using TensorFlow API we could write this as follows:
\begin{lstlisting}[language=Python,frame=single,label=prediction_tf]
predictions = tf.tensordot(features, weights, axes=1)
\end{lstlisting}
In the mathematical equation and TensorFlow, dimensions between $x_i$ and $\theta$ need to match, in order for the matrix-vector multiplication to be executed. %
Hence, the transpose operation on $x_i$ in Equation \ref{lr_prediction}. %
In SQL this part is implemented as an element-wise multiplication between columns and a group by aggregation, which implies that dimensions are irrelevant in this case. %
The translation process generates the TensorFlow snippet above from the corresponding SQL snippet. %

A SQL program is a list of SQL queries. %
We handle a specific (albeit general enough to express supervised ML algorithms) type of \texttt{select} query that is described %
in algorithm \ref{translateView}. %
We assume that there is only one numeric expression per query, although it may involve multiple nested computations, e.g. (a+b)/c.
Because we do not handle subqueries, we store the result of a query as a view and use the view name in subsequent queries when needed. %
Algorithm \ref{translateView} displays the high-level steps to translate a \texttt{create view} query to a TensorFlow command. %
The translation proceeds by applying typical steps from compiler theory. %
A SQL query is first tokenized (lexing) and parsed. %
Then an AST is produced for it. %
Based on the AST we extract the numeric expression involved in the query as well as the name of the view, and generate an equivalent TensorFlow command. %
\begin{algorithm}
\SetKwInOut{Input}{input}
\SetKwInOut{Output}{output}
\Input{a SQL view viewQuery}
\tcc{we handle create view queries of the form: \\
	CREATE VIEW \$(name) AS
	SELECT \$(columns), \$(numericExpr)
	FROM \$(tables)
	WHERE \$(joinElement)
	GROUP BY \$(groupingElement)}
\Output{a TensorFlow command tfCommand}
\tcc{and generate TensorFlow code of the form: \\
\$(name) = \$(translateNumericExpr(numericExpr))}
\BlankLine
\SetKwFunction{FtranslateView}{translateView}
  \SetKwProg{Fn}{Function}{:}{}
  \Fn{\FtranslateView{$viewQuery$}}{
  	$lexResult \leftarrow lexing(viewQuery)$\;
	$ast  \leftarrow parsing(lexResult)$\;
	$numericExpr \leftarrow getNumericExpr(ast)$\;
	$viewName \leftarrow getViewName(ast)$\;
	$return ~ code ~ viewName=translateNumericExpr(numericExpr)$\;
	
  }
\caption{Function translateView}\label{translateView}
\end{algorithm}

%\begin{algorithm}
%\SetKwInOut{Input}{input}
%\SetKwInOut{Output}{output}
%\Input{a list listSQL of SQL queries}
%\Output{a list listTF of TensorFlow commands}
%\BlankLine
%\For{$i\leftarrow 1$ \KwTo length of $listSQL$}{
%$lexResult \leftarrow lexing(listSQL[i])$\;
%$parseResult  \leftarrow parsing(lexResult)$\;
%$ast  \leftarrow generateAST(parseResult)$\;
%$arithmeticExpr \leftarrow findArithmeticExpr(ast)$\;
%$tfCommand \leftarrow generateTFCommand(arithmeticExpr)$\;
%$listTF[i] \leftarrow tfCommand$\;
%}
%\caption{Translation steps}\label{translation_algo}
%\end{algorithm}

The AST represents necessary information for translating a SQL query to functions of TensorFlow API. %
This information includes the operators that are involved in a \texttt{SELECT} numeric expression, the columns on which the operators operate, the tables where the columns came from, as well as columns involved in group by expressions. %
In order to generate the TensorFlow code, we need to match SQL numeric operators and aggregation functions with linear algebra operators. %
%However, not every relational operator can be translated to a linear one. %
%We are actually interested in numeric operators and aggregation functions, which can be matched to linear algebra operators. %
In Table \ref{tab:relational_to_linear} we present equivalences between these two categories of operators encountered in examples in this paper.

%First, each statement in the SQL code of the ML model is initially translated to an abstract syntax tree (AST). %

\begin{table}[h!]
  \begin{center}
    \caption{Equivalences between linear algebra, SQL and TensorFlow operators}
    \label{tab:relational_to_linear}
    \begin{tabular}{l|l|l}
      \textbf{Linear algebra operator} & \textbf{SQL} & \textbf{TensorFlow} \\
      \hline
      matrix addition & + & tf.add \\
      matrix subtraction & - & tf.subtract \\
      Hadamard product & $\ast$ & tf.multiply \\
      Hadamard division & / & tf.div \\
      matrix product & SUM(\_ * \_) & tf.tensordot \\
    \end{tabular}
  \end{center}
\end{table}

%In order to generate the TensorFlow code, we need to match SQL operators with TensorFlow operators. As mentioned above, not every SQL operator can be translated to a TensorFlow one. However, we are actually interested in arithmetic operators and mathematical and aggregation functions, which can be matched to TensorFlow operators. In table \ref{tab:sql_to_tf} we present equivalences between SQL and TensorFlow for the operators encountered in the example programs presented in section \ref{sec:sql_examples}.

%\nm{Can we write a pseudocode snippet for the translation process?}
%Starting with the AST of view \texttt{predictions} as depicted in figure \ref{fig:ast_products} we will now describe how the abstract syntax tree of a SQL statement is translated to TensorFlow commands (see function \texttt{generateTF} in function \ref{generate_TF_command}). %
%\begin{figure}
%  \includegraphics[width=\linewidth]{images/ast_products_larger.pdf}
%  \caption{AST of view \texttt{products}}
%  \label{fig:ast_products}
%\end{figure}

As soon as we extract the numeric expression among the \texttt{SELECT} expressions of a query, we analyze it recursively by visiting each subexpression and decomposing it to the operators and the columns on which the operators take place as described in Algorithm \ref{translateNumericExpr}. %
We apply a compositional translation where there is a mapping that preserves the structure between the SQL and TensorFlow expression.
If the expression is a column or a constant, we output a TensorFlow constant or variable name. %
Otherwise, we match it to a TensorFlow operation according to Table \ref{tab:relational_to_linear} and proceed with translating the operands of %
the expression. %
%Then, we decompose this expression to the operations and the columns on which the operations take place. %
Recall that tensors store only real values, so analyzing column projections/expressions with other types is not applicable.
%\rlw{what about 1-hot encoding for categoricals?}
%\nm{We assume that the SQL developer has created tables for 1-hot encoding representation of string columns when needed. %
%The reason for this is that it wouldn't feel right to write the objective function using string columns. %}
%When we have isolated the operator used in an expression, we match it to a TensorFlow operation as displayed in table \ref{tab:relational_to_linear}. %
%The column names mentioned in an expression as well as the tables in the \texttt{FROM} clause are used to identify the tables involved in the query. %
%Based on the names of tables, we create tensors using TensorFlow commands.
%We can consider the abstract syntax tree as a logical plan of a SQL statement. This abstract syntax tree is used to generate the TensorFlow code. The AST is not translated in its entirety to TensorFlow code, because there is not an equivalence between everything in SQL and the TensorFlow API. For example, we cannot translate joins to TensorFlow API functions. That is why we clarify that we just use the AST. The TensorFlow program can be seen as a physical plan for the initial SQL statements.
%\rlw{what can be translated? why does it matter that joins can't be translated easily?}

\begin{algorithm}
\SetKwInOut{Input}{input}
\SetKwInOut{Output}{output}
\Input{AST expr of numeric expression}
\Output{equivalent TensorFlow expression}
\SetKwFunction{FtranslateNumericExpr}{translateNumericExpr}
  \SetKwProg{Fn}{Function}{:}{}
  \Fn{\FtranslateNumericExpr{$expr$}}{
 	\If{$expr$ is a $<Constant>$}{
  		return a TensorFlow Constant
  	}
	\If{$expr$ is a $<ColumnName>$}{
  		return a TensorFlow Variable Name
  	}
	\If{$expr$ is $<Operator(Operand1, Operand2, ...)>$}{
  		return a TensorFlow $<matchTensorFlowOp(Operator)>$ with arguments $translateNumericExpr(<Operand1>), translateNumericExpr(<Operand2>)$
  	}
  }
\caption{Function translateNumericExpr}\label{translateNumericExpr}
\end{algorithm}

%\begin{table}[h!]
%  \begin{center}
%    \caption{Equivalences between SQL and TensorFlow operators}
%    \label{tab:sql_to_tf}
%    \begin{tabular}{l|c|r}
%      \textbf{SQL operator} & \textbf{TF operator} \\
%      \hline
%      SUM and * inside SUM & tf.tensordot \\
%      SUM & tf.reduce\_sum\\
%      COUNT & tf.size \\
%      AVG & tf.reduce\_mean \\
%      EXP & tf.exp \\
%      LN & tf.log \\
%      negative & tf.negative \\
%      dot & tf.tensordot \\
%      + & tf.add \\
%      - & tf.subtract \\
%      * & tf.multiply \\
%      / & tf.div \\
%      POW with 2 as exponent & tf.square
%    \end{tabular}
%  \end{center}
%\end{table}
%
%We now present how the abstract syntax trees of two SQL statements from the Linear Regression and the Logistic Regression examples above are translated to TensorFlow commands.
%
%The first one is the view \texttt{products}, which multiplies feature values with their weights and then sums over the individual products of every observation. It is basically a matrix-vector multiplication.
%
%\begin{verbatim}
%CREATE VIEW product AS
%SELECT SUM(features.v * weights.v) AS v,
%features.observationID AS observationID
%FROM features, weights
%WHERE
%features.featureName = weights.featureName
%GROUP BY observationID;
%\end{verbatim}
%
%Figure \ref{fig:ast_products} displays the abstract syntax tree of the query.
\begin{figure}
  \includegraphics[width=\linewidth]{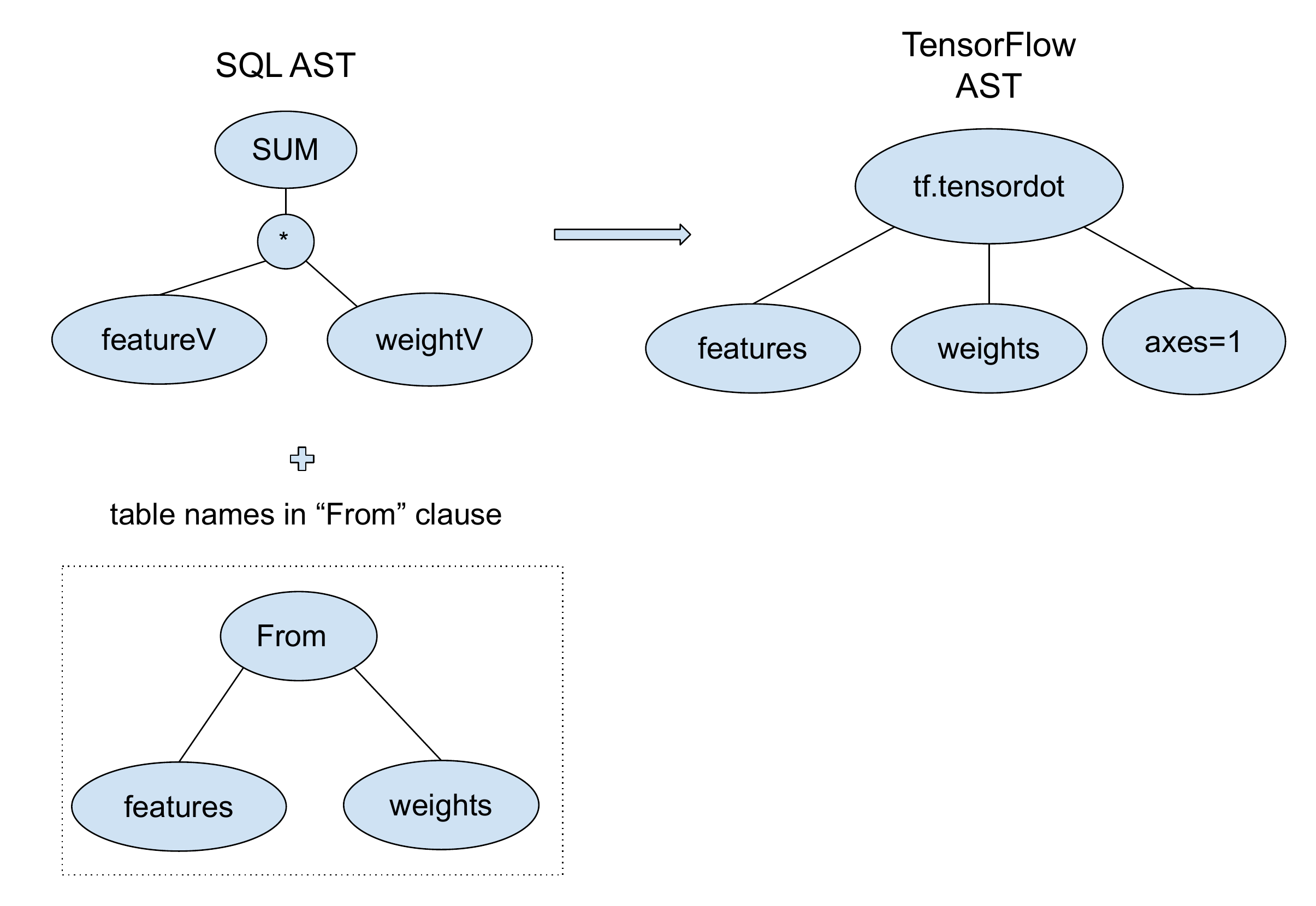}
  \caption{From SQL to TensorFlow AST of \texttt{features}-\texttt{weights} product}
  \label{fig:sql_to_tf_ast}
\end{figure}

Hence, in the select query of view \texttt{predictions} we will translate only the expression \texttt{SUM(features.v * weights.v)} to TensorFlow code. %
%\texttt{WHERE} clauses are join operations which are not applicable on tensors, whereas \texttt{GROUP BY} clauses are analyzed only in particular cases. %
In Figure \ref{fig:sql_to_tf_ast} we present the TensorFlow AST that corresponds to the SQL AST of a matrix-vector multiplication. %
Based on Table \ref{tab:relational_to_linear} we match the combination of function SUM and $\ast$ with \texttt{tf.tensordot}. %
We then scan the columns \texttt{features.v} and \texttt{weights.v}, and identify that they belong to tables \texttt{features} and \texttt{weights}. %
We use these table names for the tensors participating in the corresponding TensorFlow operation. %
Finally, the argument \texttt{axes=1} in \texttt{tf.tensordot} is used to define the axes over which the sum of products takes place. %
Since our work focuses on data that can be stored in up to two-dimensional tensors, i.e. matrices and vectors, we will always use 1 as the value of the axes argument, which is equivalent to matrix multiplication.
%For example, the expression \texttt{SUM(features.v * weights.v)} is translated to \texttt{products = tf.tensordot(features, weights, axes=1)}. The function \texttt{tf.tensordot} works with tensors of any dimensions. Parameter \texttt{axes=1} is used to define that we operate on matrices and hence the result of the function is equivalent to matrix multiplication. \texttt{WHERE} clauses are join operations which are not applicable on tensors. Finally, in this case we do not need to analyze \texttt{GROUP BY} clauses, because tf.tensordot will sum over each row anyway. We will see below how we use this information in \texttt{tf.reduce\_*} functions.

Apart from linear algebra, other mathematical operations may be encountered in machine learning algorithms. An example of this is the sigmoid function used in logistic regression, which is defined in Equation \ref{sigmoid_equation}. %
The sigmoid function is implemented in SQL as follows. %
\begin{lstlisting}[language=SQL,frame=single,label=sigmoid_sql]
CREATE VIEW sigmoid AS
  SELECT product.rowID  AS rowID,
  (1/(1+EXP(-product.v))) AS v
  FROM product;
\end{lstlisting}
In TensorFlow API we can write this using the supported sigmoid function directly
\begin{lstlisting}[language=Python,frame=single,label=sigmoid_tf]
sigmoid = tf.sigmoid(product)
\end{lstlisting}
or in a more verbose way as follows:
\begin{lstlisting}[language=Python,frame=single,label=sigmoid_tf_verbose]
sigmoid = tf.div(1,
             tf.add(1,tf.exp(tf.negative(product))))
\end{lstlisting}

The translation process is very similar to the former example, except that here we have the exponential function operating on the product between the features of each observation and the weights $\theta$. %
As a result apart from the mapping between element-wise numeric and linear algebra operators, we also need to match mathematical functions used in SQL to functions of TensorFlow API. %
All mathematical functions we have encountered so far in our SQL examples are also included in the TensorFlow API. %
We present the correspondence between the two in Table \ref{tab:math_sql_tf}.

\begin{table}[h!]
  \begin{center}
    \caption{Matching mathematical functions in SQL and TensorFlow API}
    \label{tab:math_sql_tf}
    \begin{tabular}{l|l|l}
      \textbf{Math function} & \textbf{SQL} & \textbf{TensorFlow} \\
      \hline
      sum & SUM & tf.reduce\_sum\\
      count & COUNT & tf.size \\
      average & AVG & tf.reduce\_mean \\
	 exponential & EXP & tf.exp \\
      natural logarithm & LN & tf.log \\
      negative number & - & tf.negative \\
      square & POW(\_, 2) & tf.square
    \end{tabular}
  \end{center}
\end{table}

\begin{figure}
  \includegraphics[width=\linewidth]{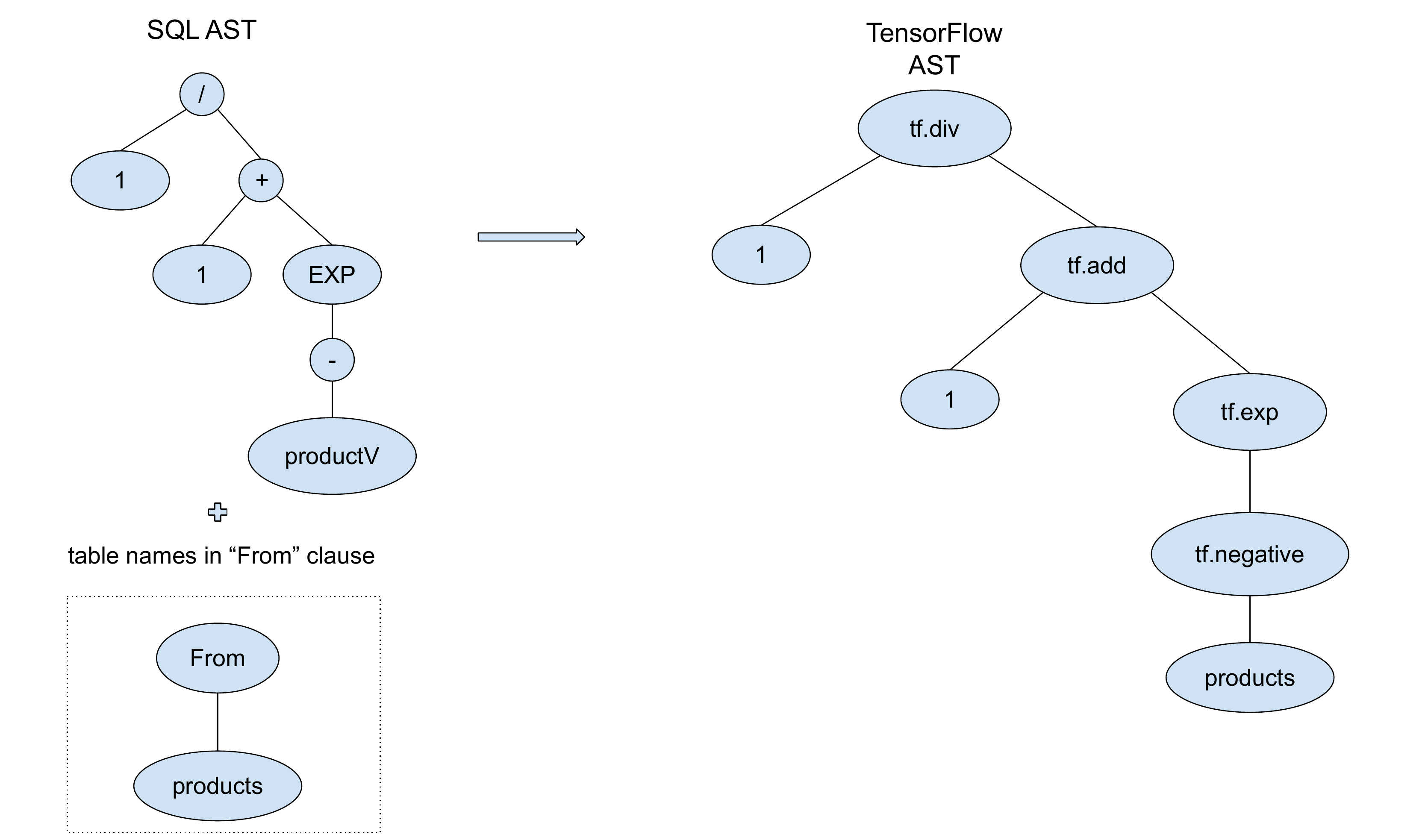}
  \caption{From SQL to TensorFlow AST of sigmoid function}
  \label{fig:sql_to_tf_ast_2}
\end{figure}

Assuming that the products between features and weights has been already defined in a different query and based on the SQL AST in Figure \ref{fig:sql_to_tf_ast_2}, we identify four operators, /, +, EXP and - in the numeric \texttt{select} expression of view \texttt{sigmoid}. %
Each of them will be matched to a TensorFlow operator forming the AST at the right of Figure \ref{fig:sql_to_tf_ast_2}. %
More specifically - is matched with \texttt{tf.negative}. %
We treat it as a different case from \texttt{tf.subtract}, when it is applied to a single constant / column. %
EXP is matched with \texttt{tf.exp}, whereas / and + are matched with \texttt{tf.div} and \texttt{tf.add} respectively. %
We combine all operators to a single TensorFlow command by nesting them. %
The generated TensorFlow command will be
\begin{lstlisting}
sigmoid = tf.div(1, tf.add(1,
       tf.exp(tf.negative(product)))).
\end{lstlisting}

%\rlw{any way to get rid of the 'to\_double'?}
%\nm{The problem here is that writing just 1 is considered as a float32 value, whereas the rest may be defined as float64. %
% So tf.to\_double converts 1 to float64. %
% I think that we can remove it, as it is an implementation detail that does not add value to the comprehension of the example.}

%The second query implements the logistic loss function used in Logistic Regression, whose abstract syntax tree is displayed in figure \ref{fig:ast_loss}.
%
%\begin{verbatim}
%CREATE VIEW loss AS
%SELECT (-1)*SUM((first_term.v + second_term.v))
%AS v
%FROM first_term, second_term
%WHERE
%first_term.observationID = second_term.observationID ;
%\end{verbatim}
%
%\begin{figure}
%  \includegraphics[width=\linewidth]{images/ast_logistic_loss.pdf}
%  \caption{AST of view named "loss"}
%  \label{fig:ast_loss}
%\end{figure}
%
%In this \texttt{SELECT} expression we identify four operators, -, SUM, * and +. Each of them will be matched to a TensorFlow operator. Specifically - is matched with \texttt{tf.negative}. We treat it as a different case from \texttt{tf.subtract}, when it is applied to a single constant / column. SUM is matched with \texttt{tf.reduce\_sum}, whereas * and + are matched with \texttt{tf.multiply} and \texttt{tf.add} respectively. We combine all operators to a single TensorFlow command by nesting them. The generated TensorFlow command will be \texttt{loss = tf.negative(tf.reduce\_sum(tf.add(firstTerm, secondTerm)))}

Finally, when we have a \texttt{group by} clause combined with an aggregation function in SQL, such as SUM or AVG, we need to analyze it, in order to extract necessary information regarding the parameters of the matched reduce operation in TensorFlow. %
In reduce operations in TensorFlow, the user also has to specify the dimensions to reduce over, i.e. %
the \texttt{axis} parameter. %
If the \texttt{group by} in the SQL statement is on a primary key, then we need to reduce across columns and the value of \texttt{axis} parameter is 1. %
If the \texttt{group by} column is not a key, then we need to reduce across rows and the \texttt{axis} parameter is 0. %
According to TensorFlow API, the \texttt{axis} parameter takes values in the range [-rank(input\_tensor), rank(input\_tensor)], but as we limit our translation method to matrices and vectors, we do not deal with higher dimensional tensors of rank greater than 2. %
In case there is no \texttt{group by} clause, that means that we give the value \texttt{None} to the \texttt{axis} parameter, and the reduction happens across all dimensions, as for example in \texttt{objective} query of logistic regression in Listing \ref{logistic_loss_sql}.

To provide a translation, apart from the SQL code, we also request from the user a few information hints. %
Those hints can be provided in the form of command line arguments or in a configuration file. %
More specifically, the names and dimensions of the following tables are needed: %

\begin{itemize}
\item Tables that store \emph{features}, e.g. \texttt{features}, %
\item Columns in the aforementioned tables that store \emph{names of the features}, e.g. \texttt{featureName} in table \texttt{features}, %
\item Tables that store \emph{weights} of the ML model, which are initially empty and will be populated once the gradient descent algorithm has computed the optimal values for them, e.g. \texttt{weights}, %
\item The \emph{dimensions} of the tables that store weights of the ML model, which can be derived from the cardinalities of all columns in the table except from the one storing the value of the weight and whose type will be double, e.g. [2]. The number of weights should also be equal to the number of features.%
\item The table that stores \emph{target} values of training observations, e.g. \texttt{targets}. %
\end{itemize}

Based on these tables, we are able to determine which tensors will be defined as constants or variables at the TensorFlow side, as well as provide their dimensions. % 
Finally, the following information are necessary for the training loop and the connection to the database. %
\begin{itemize}
\item Hyperparameters for the gradient descent algorithm: the number of iterations to run (e.g. 1000) and the learning step (e.g. 0.00001), %
\item A url and the credentials of a user to connect to the database, e.g. \texttt{mluser@localhost/items}. %
\end{itemize}

Following the translation process, the TensorFlow code that is being generated given the SQL snippet of Listing \ref{logistic_loss_sql} is displayed in Listing \ref{logistic_tf}.

\begin{lstlisting}[language=Python,frame=single,caption=Linear regression in TensorFlow,label=logistic_tf]
product=tf.tensordot(features,weights,axes=1)
sigmoid=tf.div(tf.to_double(1),tf.add(tf.to_double(1),tf.exp(tf.negative(product))))

log_sigmoid=tf.log(sigmoid)
log_1_minus=tf.log(tf.subtract(tf.to_double(1),sigmoid))

objective=tf.negative(tf.reduce_sum((tf.add(tf.multiply(targets,log_sigmoid)),(tf.multiply(tf.subtract(tf.to_double(1),targets),log_1_minus))),1))

optimizer = tf.train.GradientDescentOptimizer(3.0e-7)
train = optimizer.minimize(objective)

with tf.Session() as session:
 	session.run(tf.global_variables_initializer())
 	for step in range(1000):
 		session.run(train)
 		print("objective:", session.run(objective))
\end{lstlisting}

%When we have a reduce function in TensorFlow, we also need to specify the dimensions to reduce. In case we give no value for the \texttt{axis} parameter, the reduction will happen across all dimensions, as is the case with the query for logistic loss right above. If the query had a \texttt{GROUP BY} clause though, that would give us the information we need to determine the value of the \texttt{axis} parameter. If the group by is on a single column and that column is a primary key, then we need to reduce across columns and the value of \texttt{axis} parameter is 1. If the column is not a key, then we need to reduce across rows and the \texttt{axis} parameter is 0. According to TensorFlow API, the axis parameter takes values in the range [-rank(input\_tensor), rank(input\_tensor)], but as we limit our translation method to matrices and vectors, we do not deal with higher dimensional tensors.

\subsection{Moving data from relations to tensors}
\label{sec: moving}
So far we described how we translate SQL numeric and aggregation operations into TensorFlow operators. %
In order to execute the TensorFlow program and compute the optimal weights of the ML model, we need to feed the tensors with data. %
Data are initially stored as relations inside the database. %
So we need a mechanism to transfer the data from relations to tensors. %
However, in section \ref{sec:relational_vs_linear} we described the mismatch between the two data structures.

Let us start with possible ways of storing training data in relations. %
Suppose each observation has a number of features and a target value. %
We could store this information in three relations: \texttt{observations}, \texttt{features} and \texttt{targets}. %
The schema of these relations would be

\begin{lstlisting}[frame=single,label=schema_2]
observations(rowID: int)
features(rowID: int, name: string, v: double)
targets(rowID: int, v: double)
\end{lstlisting}

\subsubsection{Denormalization}
\label{sec:denormalization}
\begin{figure}
  \includegraphics[width=\linewidth]{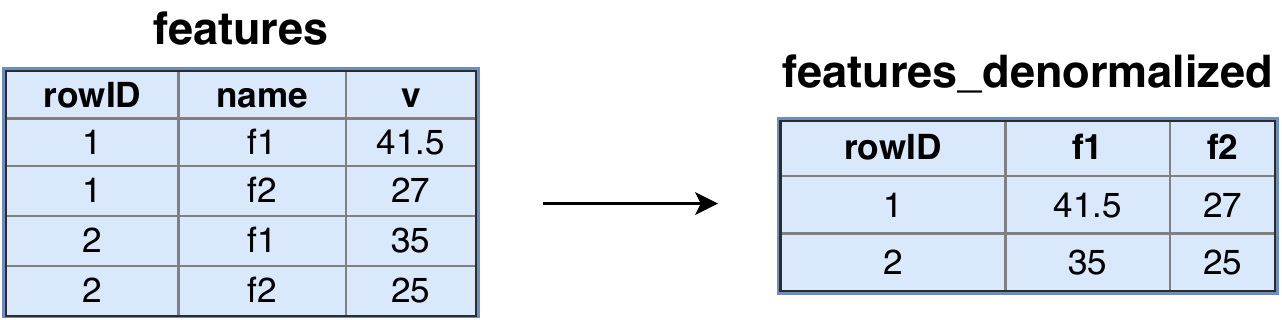}
  \caption{Denormalizing features table}
  \label{fig:features_denormalized}
\end{figure}

Given this schema we need to transfer the feature values of each observation to a matrix. %
Although relation \texttt{features} stores each feature of an observation in a different tuple, e.g., (1, "price", 3.5) and (1, "size", 20), for a matrix %
we need to consolidate all features of an observation to a single row, e.g. (3.5, 20). % 
In addition to this, we are interested only on the real values, discarding the observation IDs and names of features. %
Hence, we automatically construct a SQL query, which transforms the schema of \texttt{features} relation to a different one storing all features of an observation in a single tuple as it is displayed in Figure \ref{fig:features_denormalized}. %
The resulting schema is the following:

\begin{lstlisting}[frame=single,label=schema_2_dernomalized]
features(rowID: int, f1: double, f2: double)
\end{lstlisting}

%\rlw{illustrate the denormalize/pivot operation? what about default values?}
%\nm{Added an image for denormalization. %
% I'm not sure what you mean by default values. %}

We create a column per feature name and store only the value of the feature in it. %
The pivot query in Listing \ref{pivot_query}, which transforms rows to columns for \texttt{features} relation, is generated automatically and is added above the TensorFlow code defining the ML model, so that we can feed tensors with the appropriate data before training begins.
Essentially, a relation $R(id_i, fn_i, fv_i)$ can be formulated as a $(1..cardinality(id))\times(1..cardninality(fn))$ matrix where
\begin{multline}
M[i][j] = v,  \\ 
i \in 1..cardinality(id), j \in 1..cardninality(fn), v \in \mathbb{R}
\end{multline}
Of course, in order to be able to execute the pivot query, we also add a few lines of boilerplate code to connect to the database and run the query there. %
The result of the query is then processed with a little bit more of boilerplate code, which discards \texttt{rowID} and keeps only feature values, in order to be fed to a tensor.
\begin{lstlisting}[language=SQL,frame=single,caption=Pivot query for features,label=pivot_query]
SELECT rowID,
  SUM(CASE WHEN name='f1' THEN v ELSE 0.0 END) AS f1,
  SUM(CASE WHEN name='f2' THEN v ELSE 0.0 END) AS f2
FROM features
GROUP BY rowID;
\end{lstlisting}

Target values are fed to a vector in a similar manner. %
Because this relation stores a single tuple for each \texttt{rowID}, we do not need a pivot query. %
We rather execute a \texttt{select} query on table \texttt{targets} and keep only the target value in order to store it in a vector.

However, instead of storing all features in a single table, we could have a case where features are stored in different tables. 
Thus, we would need to join tables to gather the features of an observation. %
Suppose that we have a database storing items and their sales and we want to use it to train a ML model for future sales prediction. %
The schema of the database is

\begin{lstlisting}[frame=single,caption=Sales database schema,label=items_schema]
-- dimensions
item(itemID: string)
stores(storeID: string)
dates(dateValue: string)
-- observations
observations(itemID: string, storeID: string, dateValue: string)
-- features
familyFeat(itemID: string, family: string, v: int)
cityFeat(storeID: string, city: string, v: int)
-- targets
sales(itemID: string, storeID: string, dateValue: string, v: double)
\end{lstlisting}

\begin{figure}
  \includegraphics[width=\linewidth]{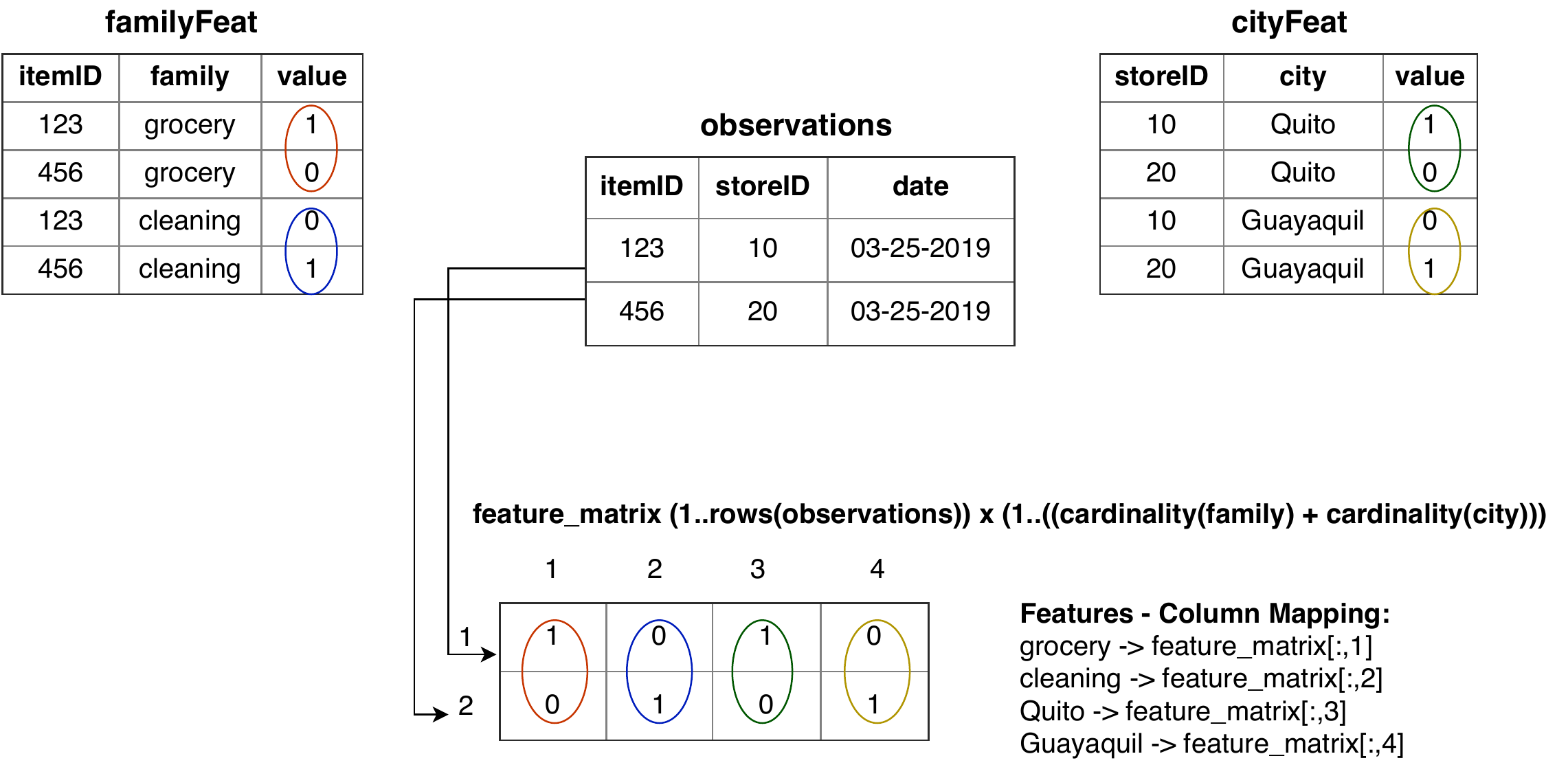}
  \caption{Mapping between features from multiple tables and the global features matrix}
  \label{fig:features_family_city}
\end{figure}

In the schema above an observation depends on three dimensions: \texttt{items}, \texttt{stores} and \texttt{dates}. %
The same holds for \texttt{sales}. %
We can use these data to predict future sales of an item at a particular store and date. %
The features of the ML model are based on item or store characteristics and are stored in \texttt{familyFeat} and \texttt{cityFeat}. %
Because families of items and cities are strings, the user has created a one-hot encoding representation in \texttt{familyFeat} and \texttt{cityFeat} relations. %
Using this representation one can convert categorical variables into a numeric form, where each family and city value becomes a real feature whose value can be either 0 or 1. %
Hence, in order to gather the feature values for an observation, we need to join \texttt{observations} with \texttt{familyFeat} and \texttt{cityFeat}. % 
Listing \ref{pivot_query_family_city} displays the automatically generated query that joins \texttt{observations} with \texttt{familyFeat} and \texttt{cityFeat} and gathers all features of an observation in a single row. %
Figure \ref{fig:features_family_city} presents the mapping between features from different tables and columns of the matrix storing features for all observations. %
It displays the export of a table to a dense matrix, but in case of categorical features with lots of zero values %
exporting to a sparse representation, such as csr \cite{Saad03}, to save on space and time is also possible.

\begin{lstlisting}[language=SQL,frame=single,caption=Pivot query for family and city features,label=pivot_query_family_city]

SELECT 
observations.itemID, observations.storeID, observations.dateValue,
groceryValue, cleaningValue, quitoValue, guayaquilValue 
FROM observations,
(SELECT itemID,
     SUM(CASE WHEN family='grocery'  THEN v ELSE 0 END) AS groceryValue,
     SUM(CASE WHEN family='cleaning' THEN v ELSE 0 END) AS cleaningValue
FROM familyFeat GROUP BY itemID) 
AS familyFeat_temp,
(SELECT storeID,
     SUM(CASE WHEN city='Quito' THEN v ELSE 0 END) AS quitoValue,
     SUM(CASE WHEN city='Guayaquil' THEN v ELSE 0 END) AS guayaquilValue
FROM cityFeat GROUP BY storeID) 
AS cityFeat_temp
WHERE observations.itemID=familyFeat_temp.itemID
 AND observations.storeID=cityFeat_temp.storeID;

\end{lstlisting}

%Hence, despite features are stored in separate relations, they are all gathered in a single matrix at the TensorFlow side.
Tables corresponding to weights of the ML model are also mapped to a single vector. %
By mapping indices of the weights vector to the initial tables and columns, we are able to store computed weights back to the database after training. %
Finally, all TensorFlow commands involving feature and weight tensors as generated by their initial representation using normalized tables are rewritten in terms of the global features and weights tables.

\subsection{Reuse of feature computations}
\label{sec:reuse}
By normalizing features across tables and connecting the observations with them using foreign keys %
we avoid redundancy in storing shared feature values multiple times. %
When gathering features to populate a global feature matrix as in the previous subsection, the repetitive nature of features can be exploited to reuse computations. %
Reuse of computations can be carried out via subqueries or precomputed tables/materialized views, both standard or widely supported techniques %
in relational databases. %

In the example schema of listing \ref{items_schema}, observations are based on three dimensions: %
\emph{items}, \emph{stores} and \emph{dates}. %
The number of observations is less or equal to $items \times stores \times dates$. %
Features that are based solely on items, stores or dates are shared across observations that concern the same item, store or date. %
For example, if we have two observations, ("item1", "store1", "date1") and ("item1", "store1", "date2"), only the features involving
date change between the two observations. %
Item and store features remain the same. %
Product family and city may be an item and store feature respectively. %
Hence, instead of recomputing family and city features for each observation, we can either compute them once using a subquery or precompute them %
and store them in a materialized view/table. %

Continuing on the example above, a query for one-hot-encoding the categorical features of family and city, and gathering all features together could be the following: %
\begin{lstlisting}[language=SQL,frame=single,caption=Query for exporting features,label=naive_query]

SELECT observations.itemID,observations.storeID,observations.dateValue,
SUM(CASE WHEN family='grocery'  THEN 1.0 ELSE 0 END) AS groceryValue,
SUM(CASE WHEN family='cleaning' THEN 1.0 ELSE 0 END) AS cleaningValue
SUM(CASE WHEN city='Quito' THEN 1.0 ELSE 0 END) AS quitoValue,
SUM(CASE WHEN city='Guayaquil' THEN 1.0 ELSE 0 END) AS guayaquilValue
FROM observations, items, stores 
WHERE observations.itemID=items.itemID AND observations.storeID=stores.storeID
GROUP BY observations.itemID, observations.storeID, observations.dateValue;

\end{lstlisting}
This query naively computes the one-hot encoding representation of family and city features using \texttt{case when} expressions for every observation. %
\mysys generates a more efficient version of the query above, as is depicted in Listing \ref{pivot_query_family_city}, making use of subqueries that denormalize the categorical features of family and city for every item and store respectively. %
Then each observation is joined on \texttt{itemID} and \texttt{storeID} with corresponding tuples from \texttt{familyFeat} and \texttt{cityFeat} tables. %
The result of the subqueries can also be materialized.

Due to the statistics the database holds, it is easy to figure out that the cardinalities of items, stores and dates are smaller than the total number of %
observations and thus that some feature computations will be repeated. %
As a result, features that are based on individual dimensions of observations can be automatically precomputed by the system. %
We report evaluation results comparing time to export features with and without precomputation in section \ref{sec:experiments}.

\subsection{Implementation} The code for translation and transferring data between data structures is implemented in Haskell \footnote{https://github.com/nantiamak/sql4ml}. %
For lexing, parsing and generating the AST of SQL code we used the open source project \emph{Queryparser} \cite{queryparser}, %
also implemented in Haskell. %
Queryparser project supports three sql-dialects (Vertica, Hive and Presto), but we found it adequate for parsing at least \texttt{select} queries, \texttt{create view} and \texttt{create table} commands in MySQL and PostgreSQL, as well. %
Following the generation of AST, we developed the rest of the code for analyzing it and generating the TensorFlow program end to end. %

A translation is essentially a function from a source type to a target type. %
As such, functional languages are a good fit for developing translation processes. %
For translating SQL to TensorFlow, SQL is captured in algebraic data types (ADTs). %
The AST of a query is represented using the ADTs provided by the Queryparser project. %
We  defined extra data types to represent objects of interest particularly to the TensorFlow translation. % 
%For example the numeric expression involved in a query is treated as a Haskell data type specified by the following constructors:
%\begin{lstlisting}
%data numericExpr = 
%	Constant String 
%	|Column String 
%	|Operator String [numericExpr]
%\end{lstlisting}
%$expr$ variable in algorithm \ref{translateNumericExpr} represents an instance of this data type. %

The assembly of a TensorFlow command, i.e. the output of \texttt{translateNumericExpr} and \texttt{getViewName} in Algorithm \ref{translateView}, is carried out through functions %
walking over the AST and performing pattern matching on it. %
Various other functions are implemented in a similar way to extract group by keys, key constraints, column expressions, create table statements and generally any information needed to generate the end-to-end TensorFlow program from the AST. %
We found that features like pattern matching and tail recursion encountered in functional languages, such as Haskell, make the manipulation of tree structures
easier than how it is developed in imperative languages.

\section{Usability Benefits}
\label{sec:benefits}
Assuming the data science workflow displayed in Figure \ref{fig:ds_workflow}, we argue that the use of \mysys has a number of benefits from a usability perspective.

\subsection{Easier maintenance}
We propose the use of SQL for both feature engineering and the definition of a ML model. %
It is a common scenario to have a data engineer exporting the appropriate data from a database and a data scientist writing code for a ML model \cite{data_scientists_data_engineers_2}, \cite{data_scientists_data_engineers}. %
The first part involves data processing pipelines built out of relational operators common in SQL, Datalog and data processing APIs, whereas the second is mainly based on linear algebra operators and probabilities, and is expressed in ML framework APIs written in imperative languages. %
Expressing the entire workflow in a single language increases ease of maintenance, as it is no longer required from the users to be familiar with more than one programming paradigms. %
Moreover, declarative database languages, such as SQL, are based on high-level, standardized abstractions, which hide their physical implementation from the user. %

\subsection{Facilitating data handling} \mysys automates the move of training data from relations to tensors and removes the burden of data exports % 
from the user. %
In addition to this, processing data with SQL also facilitates a few frequent operations in machine learning. %
When working with large datasets, training usually happens on batches. %
To create randomized batches as a means of avoiding implicit bias to a ML model, training data are shuffled. %
Using a library, such as Python Pandas, on the ML framework side for shuffling requires loading the entire training set in-memory, which results in %
out-of-memory errors when data are larger than the machine's memory specs. %
This situation can be avoided if we use a RDBMS for arranging the tuples of a table in random order. %
Also, the creation of batches themselves might fail due to memory issues. %
Lately, TensorFlow \footnote{https://www.tensorflow.org/api\_docs/python/tf/data/Dataset} and PyTorch \footnote{https://pytorch.org/tutorials/beginner/data\_loading\_tutorial.html} added data structures for handling datasets incrementally, %
but if this is not provided by the library, then a typical procedure is to load data in-memory and take random samples to create each batch. %
In this case, executing SQL queries to fetch tuples per batch can provide a solution to memory issues.

\subsection{Model management}
Because in \mysys weights of a model are stored in a table, the user is able to query them. %
For example, she might be interested in examining the range of weights in order to identify large values using a SQL query similar to the following. %

\begin{lstlisting}[language=SQL,frame=single,label=weights_query]
SELECT featureName, v FROM weights WHERE v > $(threshold);
 
 -- or
 
SELECT featureName, v FROM weights ORDER BY v;
\end{lstlisting}

In ML frameworks models are saved to files, whose extension is not standard. %
For example TensorFlow uses the .ckpt and .data extensions, whereas PyTorch uses .pt or .pth files. %
Each API provides its own functions to access parts of these files, though each has its own particularities with which the user needs to get familiar. %
%\begin{lstlisting}[language=Python,frame=single,label=weights_tf]
%vars = tf.trainable_variables()
%vars\_values = sess.run(vars)
%for var, val in zip(vars, vars\_values):
%    print(var.name, val)
%\end{lstlisting}
We argue that declarative database languages provide a standard and higher-level way to query data and as such it can be used to manage ML models, too. %
We could also store metadata regarding features and weights in tables. %
One such example is to add columns or tables keeping track of different versions of a model built from different features subsets.

\subsection{Interoperability}
\mysys generates valid and readable TensorFlow/Python code. %
It can be combined with other code in Python, which is a popular language used among the data science community. %
In addition to this, TensorFlow code can run on GPUs, providing additional hardware options for training outside the database. %

\section{Experiments}
\label{sec:experiments}
%As a first experiment I think we should check how much time the translation from SQL to TensorFlow takes and compare it with exporting data to a csv file. %
%Given that the translation is automated, it is still a win even if it takes a little more time than exporting, as to export data in the appropriate format the user needs to write an ETL script, which is considerable work.
We conducted two types of experiments: (1) the time translation takes to generate the entire TensorFlow program for three models %
defined in SQL code and various feature sets, %
(2) the time difference between exporting training observations with and without precomputed features.
%(2) the correctness of the generated code by comparing the loss function between the generated and the handwritten version of the ML model in TensorFlow. %

\textbf{Experimental setup} Experiments ran on a machine with Intel Core i7-7700 3.6 GHz, 8 cores, 16GB RAM and Ubuntu 16.04. %
TensorFlow code ran on CPUs, unless stated otherwise. %
We use PostgreSQL as an RDBMS and whenever TensorFlow API is mentioned, it refers to the Python API.

\textbf{Datasets} Two datasets are used throughout the experiments: Boston Housing \cite{Harrison78}, \cite{bostonHousing} and Favorita \cite{favorita}. %
%epsilon \cite{Yuan12}
Boston Housing is a small dataset including characteristics, such as per capita crime rate, pupil-teacher ratio, for suburbs in the Boston area and the median value of owner-occupied homes. % 
%The epsilon dataset is a synthetic dataset that can be used for binary classification as its observations are annotated with either 1 or -1. %
The Favorita dataset is a real public dataset consisting of millions of daily sales of products from different stores of Favorita grocery chain. %
Dataset characteristics are displayed in Table \ref{tab:datasets}.

\begin{table*}
  \centering
    \caption{Datasets}
    \label{tab:datasets}
    \begin{tabular}{|p{3cm}|p{2cm}|p{3cm}|p{3cm}|p{3cm}|}
     \hline
      \textbf{Dataset} & \textbf{Observations} & \textbf{Total Features} &  \textbf{Numeric features} &  \textbf{Categorical features}\\
      \hline
      Boston Housing & 506 & 13 & 13 & 0 \\
      \hline
      %epsilon & 400000 & 2000 & 2000 & 0 \\
      %\hline
      Favorita & 125497040 & 12 & 4 & 8 \\
      \hline
    \end{tabular}
\end{table*}

\subsection{Translation time from SQL to TensorFlow}
In Tables \ref{tab:translation_time} and \ref{tab:translation_time_normalized} we measure the time to translate the original SQL code provided by the user to TensorFlow code. %
For every translation we report average wall-clock time after five runs. %
The generated TensorFlow code includes all steps of the workflow, end-to-end, i.e. both code for exporting/importing data and defining the ML model. %
In Table \ref{tab:translation_time} we provide translation time from SQL implementations of three models, linear regression, Factorization Machines and logistic regression having all %
features in a single table. %
We also report lines of SQL code, as well as lines of generated and handwritten TensorFlow code. %
SQL code includes both the select queries for defining the ML model and the commands for creating the tables used in its definition. %
We assume that queries are formatted reasonably in multiple lines, with select expressions, \texttt{from}, \texttt{where} and \texttt{group by} %
clauses put in separate lines. %
Features for training these models are based on the Boston Housing dataset. %
%Results of these translations are displayed in table \ref{tab:translation_time}.

We also report translation time from a linear regression model operating on different sets of normalized features, i.e. features stored in multiple tables, %
in Table \ref{tab:translation_time_normalized}. %
Essentially, this latter experiment refers to the second case of denormalization described in section \ref{sec: moving}, %
where more rewriting steps are executed by \mysys. % 
We focus on how translation time is affected by increasing the number of involved features. %
In Table \ref{tab:translation_time_normalized} the sets of features are based on Favorita dataset. %
Product and store characteristics, such as the family of a product and the location of a store, can be used as categorical features for training a model. %
In Table \ref{tab:translation_time_normalized} we assume that categorical features are converted to a one-hot encoding representation, where each value of a categorical feature becomes a numeric feature. %

\begin{table*}
  \centering
    \caption{Time to translate SQL to TensorFlow code - all features stored in a single table}
    \label{tab:translation_time}
    \begin{tabular}{|p{4cm}|p{2cm}|p{3cm}|p{3cm}|p{3cm}|}
     \hline
      \textbf{ML model} & \textbf{Lines of SQL code} & \textbf{Translation time (sec)} & \textbf{Lines of TF generated code} &  \textbf{Lines of TF handwritten code}\\
      \hline
      Linear Regression & 30 & 0.0274 & 26 & 24 \\
      \hline
      Factorization Machines & 68 & 0.065 & 36 & 34 \\
      \hline
      Logistic Regression & 42 & 0.037 & 29 & 29 \\
      \hline
    \end{tabular}
\end{table*}

\begin{table*}
  \centering
    \caption{Time to translate a SQL Linear Regression model to TF code - features stored in multiple tables}
    \label{tab:translation_time_normalized}
    \begin{tabular}{|p{5cm}|p{3cm}|p{4cm}|p{3cm}|}
     \hline
      \textbf{Feature tables} & \textbf{Total categorical features} & \textbf{Numeric 1-hot encoded features} & \textbf{Translation time (sec)} \\
      \hline
      family, city & 2 & 55 & 0.0388 \\
      \hline
      family, city, state, store type & 4 & 76 & 0.051 \\
      \hline
      family, city, state, store type, holiday type, locale, locale\_name & 7 & 109 & 0.068 \\
      \hline
      family, city, state, store type, holiday type, locale, locale\_name, item class & 8 & 446 & 0.0906 \\
      \hline
    \end{tabular}
\end{table*}

\textbf{Takeaways} Experiments show that translation time takes a few dozens millieseconds. %
Note that translation time is not affected by the number of training observations, only by the lines of SQL code and number of features, %
which affect the generation of the denormalization queries in Listings \ref{pivot_query} and \ref{pivot_query_family_city}. %
From the results in Table \ref{tab:translation_time_normalized}, we can also observe that for the same ML model, translation is more % 
time-consuming when features are normalized across tables, but the increase remains within reasonable limits.  %
The reason for this is that the translation algorithm needs to create mappings between feature/weight tables and the columns of the global %
feature/weight matrix that concentrate all values in a single place. %
Regarding the lines of code, we observe that there might exist a negligible difference between the automatically generated
and handwritten code. %
This difference is due to the fact that generated code could be more verbose as we do not support the use of subqueries in SQL code. %
For example, in the least squared error formula $\sum_{i=1}^{N}(\hat{y}_i-y_i)^2$ the automatically generated version would
compute the square of errors and their average in two operations coming from two separate SQL queries,

\begin{lstlisting}[language=Python,frame=single,label=error_tf]
squared_errors = tf.square(tf.subtract(predictions, targets))
mean_squared_error = tf.reduce_mean(squared_errors, None)
\end{lstlisting}
whereas a Python developer would probably prefer to combine these steps in a single line, e.g.
\begin{lstlisting}[language=Python,frame=single,label=mean_squared_error_tf]
mean_squared_error = tf.reduce_mean(tf.squared_difference(prediction, targets))
\end{lstlisting}
taking advantage of the \texttt{square\_difference} operation of TensorFlow API. %
Similar differences might occur in case an operation is provided by TensorFlow API but not by SQL. % 
An example is the computation of sigmoid function. %
TensorFlow provides \texttt{tf.sigmoid}, whereas in SQL the user needs to write the formula definition. %
As a result the generated code for this specific part might involve more operations than its handwritten version.

Another point where differences in code might occur is data loading. %
The automatically generated code follows the pattern of reading features and targets from two separate files/data structures,
corresponding to the two tables/columns inside the database. %
A developer though might have other options for loading data. %
For example, all data might be exported in a single file and he might use Pandas operations to split them to two
dataframes before loading them to tensors.

%we observe that when features are normalized across tables translation takes more time, % 
%but the increase in time remains within reasonable limits and is sub-linear to the number of feature tables.

\subsection{Evaluating feature precomputation}
Using the feature reuse process described in section \ref{sec:reuse}, we compare the time needed to export training data from the Favorita dataset %
with and without precomputing and materializing shared features in tables. %
Table \ref{tab:feature_precomputation} shows time performance on exporting four different sets of features. % 
The first three sets were exported on 80000000 observations. % 
Only the last one was exported on 13870445 observations as holiday features of the Favorita dataset do not apply for most observations. %
Favorita observations are based on three dimensions: items, stores and dates. %
There are 4100 items, 54 stores and 1684 dates. %
When precomputation is used, we compute and store the values of each feature in a table, which we join with every observation during exporting. %
The naive version of the query computes every feature for every observation (see Listing \ref{naive_query}) ignoring the fact that some features
are shared among observations. %

\begin{table*}
  \centering
    \caption{Time (sec) to export training data with and without feature precomputation}
    \label{tab:feature_precomputation}
    \begin{tabular}{|P{2cm}|P{2cm}|P{3cm}|P{3cm}|P{3cm}|}
    \hline
      \textbf{Observations} & \textbf{Features} & \textbf{Time without precomputation} & \textbf{Time with precomputation} &  \textbf{Precomputation time}\\
      \hline
      \multirow{3}{*}{80000000} & 1 & 602.7 & 363.7 & 0.12 \\ \cline{2-5} 
       & 2 & 979.2 & 551.63 & 0.35 \\  \cline{2-5}
       & 4 & 1331.7 & 630 & 0.475 \\ \hline
       13870445 & 8 & 355 & 122 & 0.754 \\  \hline
    \end{tabular}
\end{table*}

Time performance results indicate that feature precomputation reduced exporting time on the Favorita dataset by about 50\% while the time needed to compute and %
store features in tables remains low. %

\subsection{Correctness of generated code}
\begin{figure}
    \centering
    \begin{minipage}{.5\linewidth}
        \centering
        \includegraphics[width=\linewidth]{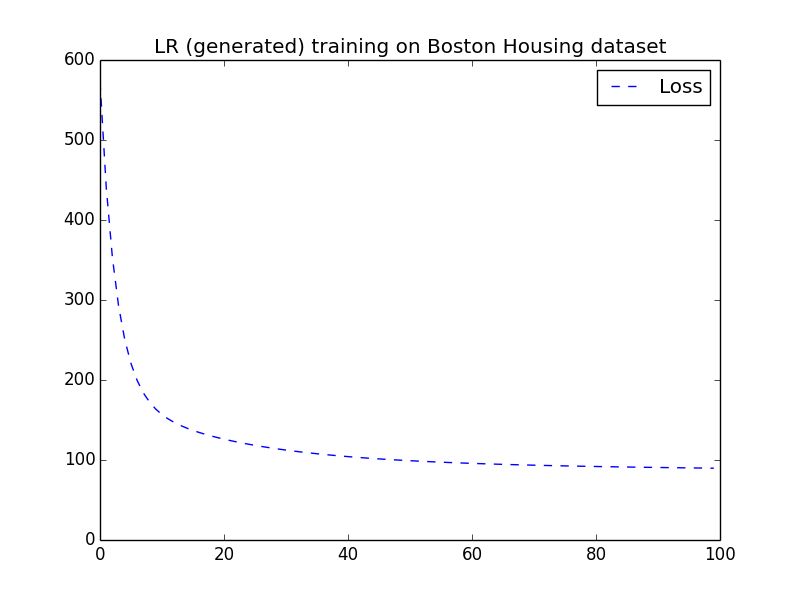} % first figure itself
    \end{minipage}\hfill
    \begin{minipage}{.5\linewidth}
        \centering
        \includegraphics[width=\linewidth]{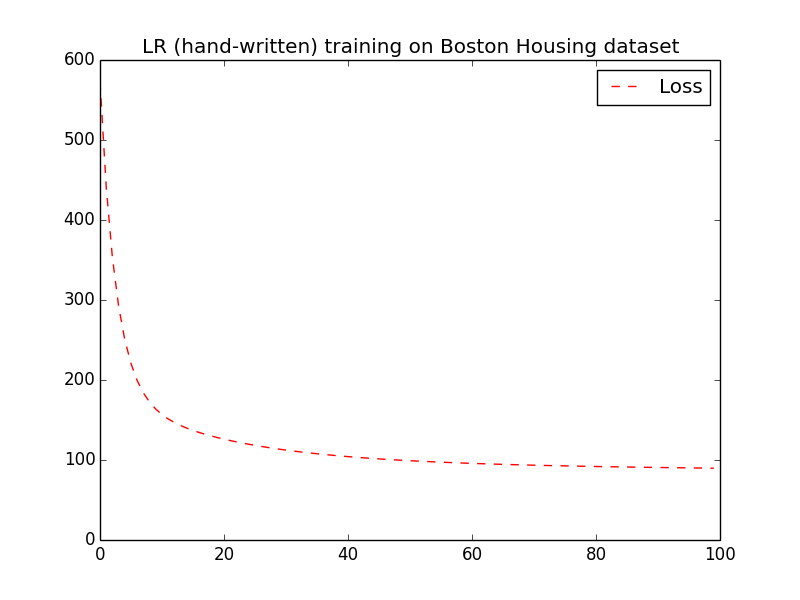} % second figure itself
    \end{minipage}
    \caption{Loss of linear regression from both generated (left) and handwritten (right) code on Boston Housing dataset}
     \label{fig:loss_lr}
\end{figure}

To check correctness of the generated code, we ran experiments with two ML models on the Boston Housing dataset of Table \ref{tab:datasets} and the epsilon dataset \cite{Yuan12}, and compared the loss function between their generated and handwritten versions. %.
The epsilon dataset is a synthetic dataset of 400000 observations with 2000 features that can be used for binary classification as its observations are annotated with either 1 or -1. % 
%Both experiments use MySQL database and TensorFlow Python API.
We trained a linear regression model on Boston Housing data that predicts the median value of owner-occupied homes using the rest of the characteristics as features. %
The learning rate used in gradient descent is 0.000003 and we ran 10000 iterations. 
The loss function is mean squared error. %
Figure \ref{fig:loss_lr} displays the loss function across iterations from both the handwritten and the generated code of the linear regression model. %
We observe that both curves follow the same decreasing pattern.

Next we provide results from two logistic regression models on the epsilon dataset. %
Due to a hard size limit of 2GB in protocol buffer \footnote{https://developers.google.com/protocol-buffers/}, TensorFlow cannot load the entire epsilon dataset at once. %
For this reason in order to train a logistic regression model, we need to perform mini-batching on the data, where each gradient descent iteration runs on a single mini-batch instead of the entire dataset. %

On the SQL side, the synthetic nature of epsilon dataset is not well-suited to the relational structure. %
All features are real values with no specific meaning. %
For our experiment we used a highly denormalized schema with 4 tables, each storing 500 features, and a table for the labels, as displayed below.% 

\begin{lstlisting}[language=SQL,frame=single,label=epsilon_schema]

features1(rowID: int, f1: doulbe, f2: double ... f500: double)
features2(rowID: int, f501: doulbe, f2: double ... f1000: double)
features3(rowID: int, f1001: doulbe, f2: double ... f1500: double)
features4(rowID: int, f1501: doulbe, f2: double ... f2000: double)
labels(rowID: int, v: int)

\end{lstlisting}

\begin{figure}
    \centering
    \begin{minipage}{.5\linewidth}
        \centering
        \includegraphics[width=\linewidth]{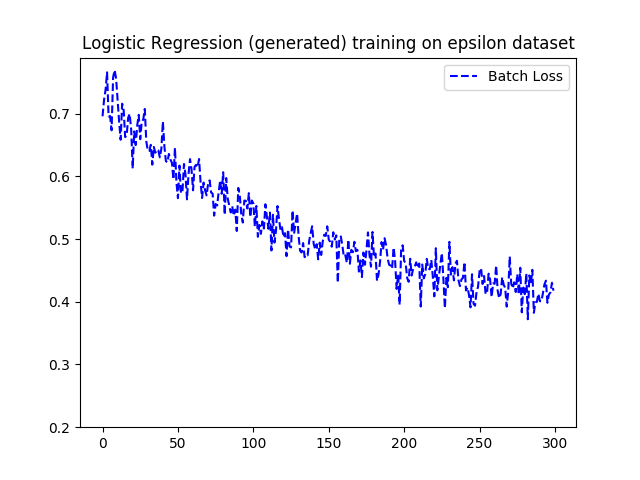} % first figure itself
    \end{minipage}\hfill
    \begin{minipage}{.5\linewidth}
        \centering
        \includegraphics[width=\linewidth]{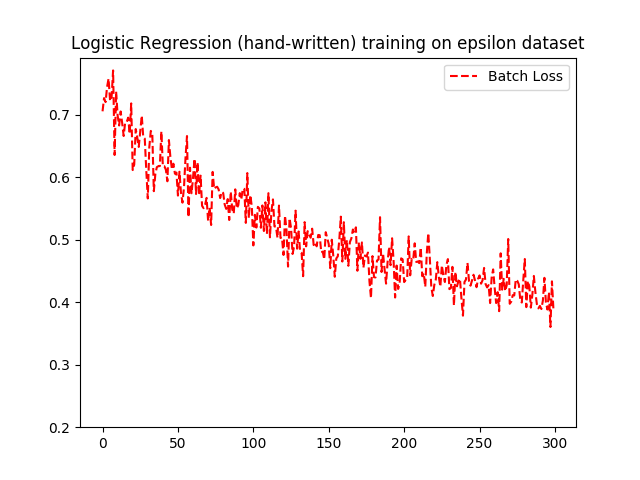} % second figure itself
    \end{minipage}
    \caption{Loss of logistic regression from both generated (left) and handwritten (right) code on epsilon dataset}
     \label{fig:loss_logistic}
\end{figure}

A query that joins all 4 tables and projects 2000 columns in order to gather all features together ran out of memory on a machine with 16GB. %
However, because TensorFlow  poses a 2GB limit due to protocol buffer anyway, we can load the dataset in two halves using two SQL queries and create mini-batches from the first half at the beginning and the second half later on. %
We compare the loss function of the generated TensorFlow code using the above configuration with a handwritten version that loads the entire dataset from a file to a NumPy array, which is then divided into mini-batches at the TensorFlow side.
For this experiment the learning rate in gradient descent was set to 0.01. % 
We ran 300 epochs and set batch size to 200. % 
Each epoch contained 2000 iterations to cover the entire training set. % 
The loss function is logistic loss. %
Figure \ref{fig:loss_logistic} displays the loss function across iterations from both versions of the logistic regression model. %
Due to the stochastic nature of mini-batching, there are small fluctuations between the two plots. % 
However, overall the pattern is very similar.

\section{Related work}
\label{sec:related}
The line of work in \mysys is mainly relevant to the following three directions. %

\textbf{In-database machine learning} There have already been efforts to bring machine learning inside the database. %
MADlib \cite{Hellerstein12} is a library of in-database methods for machine learning and data analysis. %
It provides SQL-based ML algorithms, which run on PostgreSQL \cite{postgres} or Greenplum database \cite{greenplum}. % 
SQL operators are combined with user defined functions (UDF) in Python and C++ implementing ML algorithms for popular data analysis tasks, such as classification, clustering and regression. %
Bismarck \cite{Feng12} describes a unified architecture for integrating analytics techniques as user-defined aggregates (UDAs) to RDBMS.
Similarly, Big Query \cite{BigQuery}, which is based on the Dremel technology \cite{Melnik11}, supports a set of ML models that can be called in SQL queries. %
In systems of this flavor, the user can select from a predefined collection of available algorithms. %
Moreover, code inside UDFs/UDAs remains a black box and is not optimized by the database system. %
In our approach we leave the user define the objective function of the algorithm, in order to provide more flexibility on the ML models she can develop. %
%Also, the code the user writes is standard SQL and can be optimized by the database just like any other SQL query. %

A second direction regarding in-database learning aims at mingling the dataset construction with the learning phase and accelerating the latter by %
exploiting the relational structure of the data. %
F \cite{Schleich16}, \cite{Kumar15} and AC/DC \cite{Khamis18} operate on normalized data and rewrite objective functions of ML algorithms by pushing the involved aggregations through joins. %
This idea of decomposing the computations of ML algorithms comes under the umbrella of factorized ML. %  
The purpose of these works is to showcase that given the appropriate optimizations a RDBMS can efficiently perform ML computations and the need for %
denormalization of the data can be avoided. %
Initial efforts \cite{Schleich16}, \cite{Kumar15} in this area required manual rewriting of each ML algorithm to its factorized version and because of that suffered from low generalizability. %
More recent work \cite{Chen17}, \cite{Schleich19} aims to provide a more systematic approach to automatic rewriting of ML computations to their factorized version. %
Aspects of this line of work are orthogonal to \mysys. %
For example, since our workflow also starts in the database the optimization with functional dependencies presented in \cite{AboKhamis18} can also be exploited by our translation method in order to reduce the weights of a model. %
Another example is the use of lazy joins for more efficient denormalization of data as described in Morpheus \cite{Li18}. %
We argue though that by interoperating with existing ML frameworks, we gain in usability and reduce development overhead, as we can exploit helpful features, such as automatic differentiation and mathematical optimization algorithms.

Finally, \cite{Luo17} propose an extension on SQL to support matrices/vectors and a set of linear algebra operators, whereas \cite{Jankov19} presents optimizations on executing recursion and large query plans on RDBMS, which can make them suitable for distributed machine learning. %
In \mysys we do not require any changes on the RDBMS or the machine learning framework. %
We rather propose a translation method that gets standard SQL code as input and generates valid TensorFlow/Python code.

\textbf{Machine learning frameworks} SystemML \cite{Boehm16}, TensorFlow \cite{abadi16}, PyTorch \cite{Paszke17}, Mahout Samsara \cite{Mahout} and BUDS \cite{Gao17} provide domain specific languages (DSLs) and APIs that support linear algebra operations and data structures, probability distribution and/or deep learning functions, as well as useful ML-centric features, such as automatic differentation. %
These systems are more than ML libraries as they apply algebraic rewrites and operator fusion \cite{Elgamal17}, \cite{Boehm18} to optimize users' code. %
They employ an analogy of logical and physical plans similar to query optimization in database systems.
The user essentially constructs a graph defining dependencies between operators and operands whose execution order is chosen by the system. %
Because of the emphasis on linear algebra, ML frameworks operate on denormalized data. %
\mysys targets SQL users who work with relational data and would like to perform more advanced analytics using machine learning.

\textbf{Mathematical optimization on relational data}
To the best of our knowledge, MLog \cite{Li17} and SolverBlox \cite{Klabjan18}, \cite{Makrynioti18} of the LogicBlox database \cite{Aref15} are the closest systems to \mysys. %
Both systems model ML algorithms as mathematical optimization problems using queries that find the optimal values that minimize/maximize an %
objective function defined either in Datalog or a tensor-based declarative language similar to SQL. %
MLog translates the user's code at first in Datalog and then in TensorFlow, which finally computes the optimal weights for the objective function. %
Similarly, SolverBlox, currently supporting only linear programming, translates Datalog programs to an appropriate format consumed by a linear programming solver. %
In \mysys we do not propose a new language for modeling ML algorithms, but rather we use SQL, an already popular declarative language. %
Regarding the supported ML algorithms, we expand outside linear programming as MLog does, but describe in detail a translation method directly from SQL to TensorFlow API without using any other languages for intermediate representations. % 

% Talk about these in a section about related work.
%We are aware that there have been efforts before to bring everything together in a single place. %
%These efforts concentrate on two directions. %
%They either implement machine learning algorithms inside the database \cite{Hellerstein12}, \cite{Luo17} or add relational operators to other data processing frameworks, such as Spark \cite{Zaharia10} or Python Pandas \footnote{https://pandas.pydata.org/}. %
%We argue that these approaches are sub-optimal, as they try to reinvent the wheel and do not incorporate years of expertise in providing efficient solutions for relational or linear algebra. %

\section{Conclusion and Future Work}
\label{sec:conclusion}
We presented \mysys, an integrated approach for data scientists to work entirely in SQL, while using a ML framework as an execution engine for ML computations. %
The approach of \mysys automates and facilitates tedious parts of the current fragmented data science workflow. 
We developed a prototype and showcased our translation method from SQL to TensorFlow API code on well-known ML models. % 
Evaluation results demonstrate that this translation is completed in little time . %
%and validates the correctness of the generated code. %

As future work we would like to investigate further the use of \mysys in defining and training neural networks,  such as Recurrent Neural Networks (RNN), as well as the adaptation %
of the system to unsupervised ML models. %
Another very interesting line of work is the study of query/code optimization techniques whose application could result in more efficient code on %
ML framework side. %
Recent work on in-database machine learning \cite{AboKhamis18} discusses the use of functional dependencies in reducing the dimensionality of ML models. %
Such techniques could prove useful in our approach, as well, as they could produce lower-dimensional ML models and accelerate training, %
even if this is executed outside the database. 

\bibliographystyle{abbrv}
\bibliography{sql4ml_bibliography}  % vldb_sample.bib is the name of the Bibliography in this case

%\subsection{References}
%Generated by bibtex from your ~.bib file.  Run latex,
%then bibtex, then latex twice (to resolve references).
%
%%APPENDIX is optional.
%% ****************** APPENDIX **************************************
%% Example of an appendix; typically would start on a new page
%%pagebreak
%
%\begin{appendix}
%You can use an appendix for optional proofs or details of your evaluation which are not absolutely necessary to the core understanding of your paper. 
%
%\section{Final Thoughts on Good Layout}
%Please use readable font sizes in the figures and graphs. Avoid tempering with the correct border values, and the spacing (and format) of both text and captions of the PVLDB format (e.g. captions are bold).
%
%At the end, please check for an overall pleasant layout, e.g. by ensuring a readable and logical positioning of any floating figures and tables. Please also check for any line overflows, which are only allowed in extraordinary circumstances (such as wide formulas or URLs where a line wrap would be counterintuitive).
%
%Use the \texttt{balance} package together with a \texttt{\char'134 balance} command at the end of your document to ensure that the last page has balanced (i.e. same length) columns.
%
%\end{appendix}

\end{document}